\documentclass[twocolumn,prb,superscriptaddress,floatfix,showpacs]{revtex4}
\usepackage{amsmath,amssymb}
\usepackage{graphicx}
\usepackage{dcolumn}
\usepackage{bm}

\newcommand{\AH}{\text{AH}}
\newcommand{\band}{\text{band}}
\newcommand{\bath}{\text{bath}}
\newcommand{\bk}{\mathbf{k}}
\newcommand{\bq}{\mathbf{q}}
\newcommand{\CCBFA}{\text{CCBFA}}
\newcommand{\CK}{\text{CK}}
\newcommand{\eb}{\textit{e-b}}
\newcommand{\eff}{\text{eff}}
\newcommand{\FI}{\text{FI}}
\renewcommand{\H}{\hat{H}}
\newcommand{\imp}{\text{imp}}
\newcommand{\impband}{\text{imp-band}}
\newcommand{\impbath}{\text{imp-bath}}
\newcommand{\K}{\text{K}}
\newcommand{\LC}{\text{LC}}
\newcommand{\LM}{\text{LM}}
\newcommand{\loc}{\text{loc}}
\newcommand{\n}{\hat{n}}
\newcommand{\NRG}{\text{NRG}}
\newcommand{\pdag}{\phantom{\dag}}
\newcommand{\SBM}{\text{SBM}}
\newcommand{\SC}{\text{SC}}
\newcommand{\sgn}{\text{sgn}}
\newcommand{\ZH}{\text{ZH}}

\begin{document}
\title{Quantum phase transitions in a charge-coupled Bose-Fermi Anderson model}

\author{Mengxing Cheng}
\email{mxcheng@phys.ufl.edu}
\affiliation{Department of Physics, University of Florida,
Gainesville, Florida 32611-8440, USA}

\author{Matthew T.\ Glossop}
\affiliation{Physics and Astronomy Department, Rice University,
6100 Main Street, Houston, Texas 77005, USA}

\author{Kevin Ingersent}
\affiliation{Department of Physics, University of Florida,
Gainesville, Florida 32611-8440, USA}

\date{\today}

\begin{abstract}
We study the competition between Kondo physics and dissipation within an
Anderson model of a magnetic impurity level that hybridizes with a metallic
host and is also coupled, via the impurity charge, to the displacement of a
bosonic bath having a spectral density proportional to $\omega^s$. As the
impurity-bath coupling increases from zero, the effective Coulomb interaction
between two electrons in the impurity level is progressively renormalized
from its repulsive bare value until it eventually becomes attractive. For
weak hybridization, this renormalization in turn produces a crossover from a
conventional, spin-sector Kondo effect to a charge Kondo effect. At
particle-hole symmetry, and for sub-Ohmic bath exponents $0<s<1$, further
increase in the impurity-bath coupling results in a continuous,
zero-temperature transition to a broken-symmetry phase in which the
ground-state impurity occupancy $\n_d$ acquires an expectation value
$\langle \n_d\rangle_0 \ne 1$.
The response of the impurity occupancy to a locally applied electric potential
features the hyperscaling of critical exponents and $\omega/T$ scaling that
are expected at an interacting critical point. The numerical values of the
critical exponents suggest that the transition lies in the same universality
class as that of the sub-Ohmic spin-boson model. For the Ohmic case $s=1$, the
transition is instead of Kosterlitz-Thouless type. Away from particle-hole
symmetry, the quantum phase transition is replaced by a smooth crossover, but
signatures of the symmetric quantum critical point remain in the physical
properties at elevated temperatures and/or frequencies.
\end{abstract}

\maketitle

\section{Introduction}
\label{sec:intro}

Quantum impurity models have intrigued physicists for more than half a
century.\cite{Hewson:93} In recent years, the focus has largely been on
models that exhibit quantum phase transitions (QPTs). Strictly, these are
\textit{boundary} QPTs at which only a subset of system degrees of freedom
becomes critical.\cite{Vojta:06} Boundary QPTs not only serve as prototypes
for the bulk QPTs encountered (or postulated to exist) in many strongly
correlated systems,\cite{Sondhi:97,Sachdev:99} but in certain cases they are
amenable to controlled realization in quantum-dot setups.\cite{qd-qpt-expts}

Of great current interest are dissipative quantum impurity models that describe
a dynamical local degree of freedom coupled to one or more bosonic modes
representing a frictional environment. Experiments on single-molecule
transistors\cite{set-expts} have drawn attention to transport through
nanodevices featuring electron-phonon interactions as well as local
electron-electron interactions. The essential physics of these experiments
seems to be captured in variants\cite{AH-qdot1,Cornaglia:04+05,AH-qdot2,%
Zitko:06,AH-qdot3,Dias:09} of the Anderson-Holstein model, which augments the
Anderson impurity model\cite{Anderson:61} with a Holstein coupling of the impurity
occupancy to a local (nondispersive) phonon mode. The Anderson-Holstein model
has been studied since the 1970s in connection with the phenomenon of mixed
valence,\cite{mixed-valence,Hewson:02+Jeon:03,Lee:04} and has also been adapted
to treat the effect of negative-$U$ tunneling centers on superconductivity.%
\cite{negative-U,Schuttler:88} The many theoretical approaches that have been
applied to these models have yielded general agreement that phonons serve to
reduce the effective Coulomb repulsion between electrons in the impurity level,
or even to produce an attractive net electron-electron interaction. Most
challenging has been the study of simultaneous strong Coulomb repulsion and
strong electron-phonon coupling. Here, the most robust solutions have been
provided by an extension of the numerical renormalization-group (NRG) technique,
long established as a reliable tool for tackling pure-fermionic quantum impurity
problems.\cite{Wilson:75,Krishna-murthy:80,Bulla:08} NRG studies%
\cite{Hewson:02+Jeon:03,Cornaglia:04+05,Zitko:06} have shown that in the
one-channel Anderson-Holstein model, descriptive of a single molecule coupled
symmetrically to source and drain leads, increasing the phonon coupling from
zero results in a smooth crossover from a conventional Kondo effect, involving
conduction-band screening of the impurity spin degree of freedom, to a
predominantly charge Kondo effect in which it is the impurity ``isospin'' or
deviation from half-filling that is quenched by the conduction band. However,
even for very strong electron-phonon couplings, the ground state remains a
many-body Kondo singlet and there is no QPT. By contrast, a two-channel model
describing a single-molecular transistor with a center-of-mass vibrational mode
exhibits a line of QPTs manifesting the critical physics of the two-channel
Kondo model.\cite{Dias:09}

An even greater theoretical challenge is posed by quantum impurities coupled
to dispersive bosons. A canonical example is the spin-boson model,
\cite{Leggett:87+Weiss:99} which describes tunneling within a two-state
system coupled to a bosonic bath. The model has many proposed
applications, including frictional effects on biological and chemical
reaction rates,\cite{Garg:85+Onuchic:87+Evans:95} cold atoms in a
quasi-one-dimensional optical trap,\cite{Recati:05} a quantum dot coupled
to Luttinger-liquid leads,\cite{LeHur:05} and study of entanglement
between a qubit and its environment.\cite{entanglement,LeHur:07}
In many cases, the dissipative bosonic bath can be described by a spectral
density [formally defined in Eq.\ \eqref{B} below] that is proportional to
$\omega^s$ at low frequencies $\omega$. The spin-boson model with an Ohmic
($s=1$) bath has long been known\cite{Leggett:87+Weiss:99} to exhibit a
Kosterlitz-Thouless QPT between delocalized and localized phases. The
existence of a QPT for sub-Ohmic ($0<s<1$) baths was for some years the
subject of debate.\cite{Leggett:87+Weiss:99,Spohn:85+Kehrein:96}
However, clear evidence for a continuous QPT has been provided by
the NRG,\cite{Bulla:03+05,Anders:07,LeHur:07} by perturbative expansion
in $\epsilon=s$ about the delocalized fixed point,\cite{Vojta:05}
and through exact-diagonalization calculations.\cite{Alvermann:09}

Theoretical activity has also centered on the Bose-Fermi Kondo (BFK)
model,\cite{Sengupta:00} in which an impurity spin-$\frac{1}{2}$
degree of freedom is coupled both to a fermionic band of conduction
electrons and to one or more bosonic baths. BFK models arise in the context
of unconventional heavy-fermion quantum criticality treated within extended
dynamical mean-field theory (extended DMFT) (Ref.\ \onlinecite{lcqpt})
and have also been proposed to describe quantum dots coupled either to a noisy
environment\cite{LeHur:04+Li:04+Li:05+Borda:05} or to ferromagnetic
leads.\cite{Kirchner:05+08} Studies of BFK models having different spin
rotation symmetry---SU(2), XY, or Ising---employing either
expansion\cite{Zhu:02+Zarand:02} in $\epsilon=1-s$ or the
NRG (Refs.\ \onlinecite{Glossop:05} and \onlinecite{Glossop:07}) have found continuous QPTs between
phases exhibiting the Kondo effect and localized phases in which impurity
spin flips are suppressed by the coupling to the bosonic bath(s).
For exponents $0<s<1$, most evidence suggests that the continuous QPTs of the
spin-boson and of Ising-anisotropic BFK models are equivalent. QPTs outside
the spin-boson universality class have been identified in dissipative models
featuring a pseudogap in the electronic density of states.
\cite{Chung:07+Glossop:08}

In this paper, we combine the themes outlined in the preceding paragraphs
by investigating a \textit{charge-coupled Bose-Fermi Anderson} \upshape{(}\textit{BFA}\upshape{)} \textit{model} in which
the impurity not only hybridizes with conduction-band electrons but also is
coupled, via its electron occupancy, to a bath representing acoustic phonons
or other bosonic degrees of freedom whose dispersion extends to zero energy.
The model was introduced more than 30 years ago\cite{Riseborough:77,Haldane:77a,
Haldane:77b} in connection with the mixed-valence problem. A spinless version of
the model was also discussed in the same context.\cite{Hewson:80} More recently,
very similar models have been shown to arise as effective impurity problems in
the extended DMFT for one- and two-band extended Hubbard models.\cite{Smith:99,Smith:00}
Hitherto, only limited progress has been made toward understanding the
physics of such models, and we are aware of no study of their possible QPTs.

Our NRG study of the charge-coupled BFA model with bosonic baths characterized
by exponents $0<s\le 1$ reveals a crossover with increasing electron-boson
(\eb) coupling from a spin Kondo effect to a charge Kondo effect, very similar
to that noted previously in the Anderson-Holstein model.%
\cite{Hewson:02+Jeon:03,Cornaglia:04+05,Zitko:06}
However, under conditions of strict particle-hole symmetry, further increase in
the \eb\ coupling leads to complete suppression of Kondo physics at a quantum
critical point. Beyond the critical \eb\ coupling lies a localized phase in
which charge fluctuations on the impurity site are frozen. For sub-Ohmic baths
($0<s<1$), the QPT is continuous and the numerical values of the critical
exponents describing the response of the impurity charge to a locally applied
electric potential demonstrate that the transition belongs to the same
universality class as that of the spin-boson and Ising BFK models.
For Ohmic baths (corresponding to $s=1$), the QPT is found to be of
Kosterlitz-Thouless type. Particle-hole asymmetry acts in a manner analogous
to a magnetic field at a conventional ferromagnetic ordering transition,
smearing the discontinuous change in the ground-state as a function of \eb\
coupling into a smooth crossover. Signatures of the symmetric quantum critical
point remain in the physical properties at elevated temperatures and/or
frequencies.

It is important to note that questions have been raised as to whether or not
the NRG method reliably captures the quantum critical behavior of the spin-boson
and Ising BFK models for bath exponents $0<s<\frac{1}{2}$. It is a standard
belief\cite{Sondhi:97,Sachdev:99} that the low-energy behavior near a
quantum phase transition in $d$ spatial dimensions is equivalent to that of
a classical transition in $d+z$ dimensions, where $z$ is the dynamical
exponent. In the case of the spin-boson and Ising BFK models, for which $d=0$
and $z=1$, the corresponding classical system is a one-dimensional Ising chain
with long-ranged interactions that decay for large separations $r$ like
$r^{-(1+s)}$. The Ising chain is known to possess an interacting critical point
for $\frac{1}{2}<s<1$, but to exhibit a mean-field
transition\cite{Fisher:72+Luijten:96+Luijten:97} for $0<s<\frac{1}{2}$.
By contrast, NRG studies of the spin-boson\cite{Vojta:05} and Ising
BFK (Refs.\ \onlinecite{Glossop:05} and \onlinecite{Glossop:07}) models have found non-mean-field behavior
extending over the entire range $0<s<1$, leading to a claim of breakdown of
the quantum-to-classical mapping.\cite{Vojta:05} This claim has recently
been contradicted by continuous-time Monte Carlo\cite{Winter:09} and exact
diagonalization\cite{Alvermann:09} studies. Debate is ongoing concerning the
interpretation of these various results.\cite{Winter:09,Kirchner:09+Vojta:09}
The eventual resolution of this debate may determine the validity of the
small subset of our NRG results that concerns the critical exponents of the
charge-coupled BFA model with bath exponents $0<s<\frac{1}{2}$. There is every
reason to believe that the remaining results are physically sound.

The rest of this paper is organized as follows. Section \ref{sec:model}
introduces the charge-coupled BFA Hamiltonian and summarizes the NRG method
used to solve the model. Section \ref{sec:preliminaries} contains a
preliminary analysis of the model, focusing on the bosonic renormalization
of the effective electron-electron interaction within the impurity level.
Numerical results for the symmetric model with sub-Ohmic ($0<s<1$) dissipation
are presented and interpreted in Sec.\ \ref{sec:sub-ohmic}. Section
\ref{sec:ohmic} treats the symmetric model with Ohmic ($s=1$) dissipation.
Section \ref{sec:asymm} discusses the effects of particle-hole asymmetry. The
paper's conclusions are presented in Sec.\ \ref{sec:summary}.

\section{Model and Solution Method}
\label{sec:model}

\subsection{Charge-coupled Bose-Fermi Anderson Hamiltonian and related models}
\label{subsec:model}

In this work, we investigate the charge-coupled Bose-Fermi Anderson
model described by the Hamiltonian
\begin{equation}
\label{H_CCBFA}
\H_{\CCBFA} = \H_{\imp} + \H_{\band} + \H_{\bath}
    + \H_{\impband} + \H_{\impbath},
\end{equation}
where
\begin{align}
\label{H_imp}
\H_{\imp}
&= \epsilon_d \n_d + U \n_{d\uparrow} \n_{d\downarrow}, \\
\label{H_band}
\H_{\band}
&= \sum_{\bk,\sigma} \epsilon_{\bk} \,
   c_{\bk\sigma}^{\dag} c_{\bk\sigma}^{\pdag} , \\
\label{H_bath}
\H_{\bath}
&= \sum_{\bq} \omega_{\bq} \, a_{\bq}^{\dag} a_{\bq}^{\pdag} , \\
\label{H_impband}
\H_{\impband}
&= \frac{1}{\sqrt{N_k}} \sum_{\bk,\sigma}
     \bigl( V_{\bk} c_{\bk\sigma}^{\dag} d_{\sigma}^{\pdag} +
      V_{\bk}^* d_{\sigma}^{\dag} c_{\bk\sigma}^{\pdag} \bigr), \\
\label{H_impbath}
\H_{\impbath}
&= \frac{1}{\sqrt{N_q}} \, (\n_d - 1) \sum_{\bq} \lambda_{\bq}
   \bigl( a_{\bq}^{\pdag} + a_{-\bq}^{\dag} \bigr).
\end{align}
Here, $d_{\sigma}$ annihilates an electron of spin $z$ component
$\sigma=\pm\frac{1}{2}$ (or $\sigma=\:\uparrow,\,\downarrow$) and energy
$\epsilon_d<0$ in the impurity level, $\n_{d\sigma} = d_{\sigma}^{\dag}
d_{\sigma}^{\dag}$, $\n_d = \n_{d\uparrow} + \n_{d\downarrow}$,
and $U>0$ is the Coulomb repulsion between two electrons in the impurity level.
$V_{\bk}$ is the hybridization between the impurity and a
conduction-band state of energy $\epsilon_{\bk}$ annihilated by fermionic
operator $c_{\bk\sigma}$, and $\lambda_{\bq}$ characterizes the coupling of the
impurity occupancy to bosons in an oscillator state of energy $\omega_{\bq}$
annihilated by operator $a_{\bq}$. $N_k$ is the number of unit cells in the
host metal and, hence, the number of inequivalent $\bk$ values. Correspondingly,
$N_q$ is the number of oscillators in the bath, and the number of distinct
values of $\bq$. Without loss of generality, we take $V_{\bk}$ and
$\lambda_{\bq}$ to be real and non-negative.  Throughout the paper, we drop all
factors of the reduced Planck constant $\hbar$, Boltzmann's constant $k_B$, the
impurity magnetic moment $g\mu_B$, and the electronic charge $e$.

To focus on the most interesting physics of the model, we assume a constant
hybridization $V_{\bk}=V$ and a flat conduction-band density of states (per
unit cell, per spin-$z$ orientation)
\begin{equation}
\label{rho}
\rho(\epsilon)
\equiv \frac{1}{N_k}\sum_{\bk} \delta(\epsilon - \epsilon_{\bk})
= \begin{cases}
    \rho_0=(2D)^{-1} & \mbox{for } |\epsilon|<D \\
    0 & \mbox{otherwise},
  \end{cases}
\end{equation}
defining the hybridization width $\Gamma=\pi\rho_0 V^2$. The
bosonic bath is completely specified by its spectral density,
which we take to have the pure power-law form
\begin{align}
\label{B}
B(\omega)
&\equiv \frac{\pi}{N_q} \sum_{\bq} \lambda_{\bq}^2 \,
  \delta(\omega-\omega_{\bq}) \notag \\
&= \begin{cases}
  (K_0 \lambda)^2 \Omega^{1-s}\omega^s & \mbox{for } 0<\omega<\Omega \\
  0 & \mbox{otherwise} ,
  \end{cases}
\end{align}
characterized by an upper cutoff $\Omega$, an exponent $s$ that must
satisfy $s>-1$ to ensure normalizability, and a dimensionless prefactor
$K_0\lambda$. In this paper, we present results only for the case $\Omega=D$
in which the bath and band share a common cutoff. We also adopt the
convention that $K_0$ is held constant while one varies $\lambda$, which we
term the electron-boson (\eb) coupling. It should be emphasized, though,
that the key features of the model are a nonvanishing Fermi-level density of
states $\rho(0)>0$ and the asymptotic behavior $B(\omega)\propto\omega^s$ for
$\omega\to 0$. Relaxing any or all of the remaining assumptions laid out in
this paragraph will not alter the essential physics of the model, although it
may affect nonuniversal properties, such as the locations of phase boundaries.

For many purposes, it is convenient to rewrite\cite{Krishna-murthy:80} the
impurity part of the Hamiltonian (dropping a constant term $\epsilon_d$)
\begin{equation}
\label{H_imp:delta}
\H_{\imp} = \delta_d (\n_d - 1) + \frac{U}{2} (\n_d - 1)^2 ,
\end{equation}
where $\delta_d = \epsilon_d + U/2$.
Most of the results presented below were obtained for the symmetric
model characterized by $\epsilon_d = -U/2$ or $\delta_d = 0$, for which
the impurity states $n_d = 0$ and $n_d = 2$ are degenerate in energy.
Section \ref{sec:asymm} addresses the behavior of the asymmetric model.

In any realization of $\H_{\CCBFA}$ involving coupling of acoustic phonons
to a magnetic impurity or a quantum dot, the value of the bath exponent $s$
will depend on the precise interaction mechanism. However, phase space
considerations suggest that any such system will lie in the super-Ohmic regime
$s>1$. Models closely related to $\H_{\CCBFA}$ have also been considered in the
context of extended DMFT,\cite{Smith:00,edmft} a technique for systematically
incorporating some of the spatial correlations that are omitted
from the conventional DMFT of lattice fermions.\cite{Georges:95}
Extended DMFT maps the lattice problem onto a quantum impurity problem in
which a central site interacts with both a fermionic band and one or more
bosonic baths, the latter representing fluctuating effective fields due to
interactions between different lattice sites. The charge-coupled BFA model
serves as the mapped impurity problem for various extended Hubbard models
with nonlocal density-density interactions.\cite{Smith:99,Smith:00} In these
settings, the effective bath exponent $s$ is not known \textit{a priori},
but is determined through self-consistency conditions that ensure that the
central site is representative of the lattice as a whole. The extended DMFT
treatment of other lattice models\cite{lcqpt} gives rise to exponents $0<s<1$,
and we expect this also to be the case for the extended Hubbard models.

At the Hartree-Fock level,\cite{Haldane:77a} the impurity properties of
Hamiltonian \eqref{H_CCBFA} are identical to those of the
Anderson-Holstein Hamiltonian,
\begin{equation}
\label{H_AH}
\H_{\AH} = \H_{\text{A}} + \omega_0 a^{\dag} a^{\pdag}
   + \lambda_0 (\n_d - 1) (a^{\pdag} \!\! + a^{\dag}) ,
\end{equation}
which augments the well-studied Anderson impurity model,\cite{Anderson:61}
\begin{equation}
\label{H_A}
\H_{\text{A}} = \H_{\imp}+\H_{\band}+\H_{\impband},
\end{equation}
with a Holstein coupling of the impurity charge to a single phonon mode of
energy $\omega_0$. At several points in the sections that follow, we compare
and contrast our results for $\H_{\CCBFA}$ with those obtained previously for
$\H_{\AH}$.

\subsection{Numerical renormalization-group method}
\label{subsec:NRG}

We solve the charge-coupled BFA model using the NRG method,\cite{Wilson:75,
Krishna-murthy:80,Bulla:08} as recently
extended to treat models involving both dispersive bosons and dispersive
fermions.\cite{Glossop:05,Glossop:07} The full range of conduction-band
energies $-D<\epsilon<D$ (bosonic-bath energies $0<\omega<\Omega$) is
divided into a set of logarithmic intervals bounded by
$\epsilon=\pm D\Lambda^{-k}$ ($\omega=\Omega\Lambda^{-k}$) for $k=0,1,2,...$,
where $\Lambda>1$ is the Wilson discretization parameter. The continuum of
states within each interval is replaced by a single state, namely, the
particular linear combination of band (bath) states within the interval
that enters $\H_{\impband}$ ($\H_{\impbath}$).
The discretized model is then transformed into a tight-binding form involving
two sets or orthonormalized operators:
(i) $f_{n\sigma}$ ($n=0$, 1, 2, $\ldots$) constructed as linear combinations
of all $c_{\bk\sigma}$ having $|\epsilon_{\bk}|<D\Lambda^{-n}$; and (ii) $b_m$
($m=0$, 1, 2, $\ldots$) mixing all $a_{\bq}$ such that
$0<\omega_{\bq}<\Omega\Lambda^{-m}$.
This procedure maps the last four parts of Hamiltonian \eqref{H_CCBFA} to
\begin{gather}
\label{H_band:NRG}
\H_{\band}^{\NRG}
= D \! \sum_{\substack{n=0 \\ \sigma}}^{\infty}
   \bigl[ \epsilon_n f_{n\sigma}^{\dag} f_{n\sigma}^{\pdag}
   \! + \! \tau_n \bigl( f_{n\sigma}^{\dag} f_{n-1,\sigma}^{\pdag}
   \! + \! f_{n-1,\sigma}^{\dag} f_{n\sigma}^{\pdag} \bigr) \bigr] , \\
\label{H_bath:NRG}
\H_{\bath}^{\NRG}
= \Omega \sum_{m=0}^{\infty} \bigl[ e_m b^{\dag}_m b^{\pdag}_m +
   t_m \bigl( b_m^{\dag} b_{m-1}^{\pdag}
              + b_{m-1}^{\dag} b_m^{\pdag} \bigr) \bigr] , \\
\label{H_impband:NRG}
\H_{\impband}^{\NRG}
= \sqrt{\frac{2\Gamma D}{\pi}} (f_{0\sigma}^{\dag} d_{\sigma}^{\pdag}
   + d_{\sigma}^{\dag} f_{0\sigma})^{\pdag}, \\
\label{H_impbath:NRG}
\H_{\impbath}^{\NRG}
= \frac{\Omega K_0 \lambda}{\sqrt{\pi(s+1)}} \, (\n_d - 1)
   \bigl( b_0^{\pdag} + b_0^{\dag} \bigr) .
\end{gather}
Here, $\tau_0 = t_0 = 0$, while the remaining coefficients $\epsilon_n$,
$\tau_n$, $e_m$, and $t_m$, which include all information about the
conduction-band density of states $\rho(\epsilon)$ and the bosonic spectral
density $B(\omega)$, are calculated via Lanczos recursion
relations.\cite{Glossop:07} For a particle-hole-symmetric density of
states such as that in Eq.\ \eqref{rho}, $\epsilon_n = 0$ for all $n$.

The coefficients $\tau_n$ in Eq.\ \eqref{H_band:NRG} vary for large $n$
as $D\Lambda^{-n/2}$, while $e_m$ and $t_m$ entering Eq.\ \eqref{H_bath:NRG}
vary for large $m$ as $\Omega\Lambda^{-m}$. Therefore, the problem can be
solved iteratively by diagonalization of a sequence of Hamiltonians
$\hat{H}_N$ ($N = 0$, 1, 2, $\ldots$) describing tight-binding chains
of increasing length. At iteration $N \ge 0$, Eq.\ \eqref{H_band:NRG} is
restricted to $0\le n \le N$, while Eq.\ \eqref{H_bath:NRG} is limited to
$0\le m\le M(N)$. The spirit of the NRG is to treat fermions and bosons of
the same energy scale at the same iteration. Since the bosonic coefficients
decay with site index twice as fast as the fermionic coefficients, after a
few iterations the iterative procedure requires extension of the bosonic
chain only for every second site added to the fermionic chain. In this
work, we have chosen for simplicity to work with a single high-energy cutoff
scale $D\equiv\Omega$. It is then convenient to add to the bosonic chain at
every even-numbered iteration, so that the highest-numbered bosonic site
is $M(N) = \lfloor N/2 \rfloor$, where $\lfloor x\rfloor$ is the greatest
integer less than or equal to $x$.

The NRG method relies on two additional approximations. Even for pure-fermionic
problems, it is not feasible to keep track of all the eigenstates because the
dimension of the Fock space increases rapidly as we add sites to the chains.
Therefore, only the lowest lying $N_s$ many-particle states can be retained
after each iteration. The presence of bosons adds the further complication that
the Fock space is infinite-dimensional even for a single-site chain, making
it necessary to restrict the maximum number of bosons per chain site to a
finite number $N_b$. Provided that $N_s$ and $N_b$ are chosen to be
sufficiently large (as discussed in Sec.\ \ref{subsec:NRGflows}), the NRG
solution at iteration $N$ provides a good account of the impurity contribution
to physical properties at temperatures $T$ and frequencies $\omega$ of order
$D\Lambda^{-N/2}$.

Hamiltonian \eqref{H_CCBFA} commutes with the total spin-$z$ operator
\begin{equation}
\label{S_z}
\hat{S}_z = \frac{1}{2} (\n_{d\uparrow} - \n_{d\downarrow})
    + \frac{1}{2} \sum_n \bigl( f_{n\uparrow}^{\dag} f_{n\uparrow}^{\pdag}
       - f_{n\downarrow}^{\dag} f_{n\downarrow}^{\pdag} \bigr) ,
\end{equation}
the total spin-raising operator
\begin{equation}
\label{S+}
\hat{S}_+ = d_{\uparrow}^{\dag} d_{\downarrow}^{\pdag}
    + \sum_n f_{n\uparrow}^{\dag} f_{n\downarrow}^{\pdag}
    \equiv \bigl( \hat{S}_- \bigr)^{\dag} ,
\end{equation}
and the total ``charge'' operator
\begin{equation}
\label{Q}
\hat{Q} = \n_d - 1 + \sum_n
  \bigl( f_{n\uparrow}^{\dag} f_{n\uparrow}^{\pdag}
       + f_{n\downarrow}^{\dag} f_{n\downarrow}^{\pdag} - 1 \bigr) ,
\end{equation}
which measures the deviation from half-filling of the total electron number.
One can interpret
\begin{equation}
\label{isospin}
\hat{I}_z = \frac{1}{2} \, \hat{Q}, \quad
\hat{I}_+ = - d_{\uparrow}^{\dag} d_{\downarrow}^{\dag}
   + \sum_n (-1)^n f_{n\uparrow}^{\dag} f_{n\downarrow}^{\dag}
    \equiv \bigl( \hat{I}_- \bigr)^{\dag}
\end{equation}
as the generators of an SU(2) isospin symmetry (originally dubbed
``axial charge'' in Ref.\ \onlinecite{Jones:88}).
Since $[\H_{\impbath}, \hat{I}_{\pm}]\not= 0$, the charge-coupled BFA model
does not exhibit full isospin symmetry. However, this symmetry turns out
to be recovered in the asymptotic low-energy behavior at certain
renormalization-group fixed points.

As described in Ref.\ \onlinecite{Krishna-murthy:80}, the computational effort
required for the NRG solution of a problem can be greatly reduced by taking
advantage of these conserved quantum numbers. In particular, it is possible to
obtain all physical quantities of interest while working with a reduced basis
of simultaneous eigenstates of $\hat{S}^2$, $\hat{S}_z$, and $\hat{Q}$ with
eigenvalues satisfying $S_z = S$. With one exception noted in
Sec.\ \ref{subsec:crossover}, any $N_s$ value specified below represents the
number of retained $(S,Q)$ multiplets, corresponding to a considerably larger
number of $(S,S_z,Q)$ states.

Even when advantage is taken of all conserved quantum numbers, NRG treatment of
the charge-coupled BFA model remains much more demanding than that of the
Anderson model [Eq.\ \eqref{H_A}] or the Anderson-Holstein model
[Eq.\ \eqref{H_AH}]. Being nondispersive,
the bosons in the last model enter only the atomic-limit Hamiltonian $\H_0$,
allowing solution via the standard NRG iteration procedure.
For Bose-Fermi models such as $\H_{\CCBFA}$, the need to extend a bosonic chain
as well as a fermionic one at every even-numbered iteration $N>0$, expands
the basis of $\H_N$ from $4 N_s$ states to $4 (N_b+1) N_s$ states, and multiplies
the CPU time by a factor $\sim (N_b+1)^3$. Since we typically use $N_b=8$ or 12
in our calculations, the increase in computational effort is considerable.

The choice of value for the NRG discretization parameter $\Lambda$ involves
trade-offs between discretization error (minimized by taking $\Lambda$ to
be not much greater than 1) and truncation error (reduced by working with
$\Lambda\gg 1$). Experience from other
problems\cite{Ingersent:02,Glossop:05,Glossop:07} indicates that critical
exponents can be determined very accurately using quite a large $\Lambda$.
Most of the results presented in the remaining sections of the paper were
obtained for $\Lambda=9$, with $\Lambda=3$ being employed in the calculation of
the impurity spectral function. For convenience in displaying these results, we
set $\Omega = D = 1$ and omit all factors of $\rho_0$ and $K_0$.

\section{Preliminary Analysis}
\label{sec:preliminaries}

We begin by examining the special cases in which the impurity level is
decoupled either from the conduction band or from the bosonic bath.
Understanding these cases allows us to establish some expectations for
the behavior of the full model described by Eq.\ \eqref{H_CCBFA}.

\subsection{Zero hybridization}
\label{subsec:Gamma=0}

If one sets $\Gamma=0$ in Eq.\ \eqref{H_CCBFA}, then the conduction
band completely decouples from the remaining degrees of freedom and
can be dropped from the model, leaving the zero-hybridization model
\begin{multline}
\label{H_CCBA}
\H_{\ZH} = \delta_d (\n_d-1) + \frac{U}{2} (\n_d-1)^2
  + \sum_{\bq} \omega_{\bq} a_{\bq}^{\dag} a_{\bq}^{\pdag} \\
  + \frac{1}{\sqrt{N_q}} \, (\n_d-1) \sum_{\bq} \lambda_{\bq}
    \bigl( a_{\bq}^{\pdag} + a_{-\bq}^{\dag} \bigr) .
\end{multline}
The Fock space separates into sectors of fixed impurity occupancy
($n_d=0$, 1, or 2), within each of which the Hamiltonian can be recast,
using displaced-oscillator operators
\begin{equation}
\label{a:displaced}
\bar{a}_{n_d,\bq} = a_{\bq} +
  \frac{\lambda_{\bq}}{\sqrt{N_q} \: \omega_{\bq}} (n_d-1) ,
\end{equation}
in the trivially solvable form
\begin{equation}
\label{H_CCBA:displaced}
\H_{\ZH}(n_d) = \H'_{\imp} + \sum_{\bq} \omega_{\bq}
  \bar{a}_{n_d,\bq}^{\dag} \bar{a}_{n_d,\bq}^{\pdag} ,
\end{equation}
where
\begin{equation}
\label{H'_imp}
\H'_{\imp} = \delta_d (\n_d-1) + \frac{U_{\eff}}{2} (\n_d-1)^2 .
\end{equation}
The bosons act on the impurity to reduce the Coulomb interaction
from its bare value $U$ to an effective value
\begin{equation}
\label{Ueff0}
U_{\eff}
  = U - \frac{2}{N_q} \sum_{\bq} \frac{\lambda_{\bq}^2}{\omega_{\bq}}
  = U - \frac{2}{\pi} \int_0^{\infty} \frac{B(\omega)}{\omega} \, d\omega .
\end{equation}

For the bath spectral density in Eq.\ \eqref{B} with $-1<s\le 0$,
one finds that for any nonzero \eb\ coupling $\lambda$, $U_{\eff}=-\infty$
and the singly occupied impurity states drop out of the problem.
For the remainder of this section, however, we will instead focus on bath
exponents $s>0$, for which Eqs.\ \eqref{B} and \eqref{Ueff0} give
\begin{equation}
\label{Ueff0:s>0}
U_{\eff} = U - \frac{2 (K_0 \lambda)^2}{\pi s} \, \Omega .
\end{equation}
For weak \eb\ couplings, $U_{\eff}$ is positive and the ground state
of $\H_{\ZH}$ lies in the sector $n_d=1$ where the impurity has a spin $z$
component $\pm \frac{1}{2}$. However, $U_{\eff}$ is driven negative for sufficiently
large $\lambda$, placing the ground state in the sector $n_d=0$ or $n_d=2$
where the impurity is spinless but has a charge (relative to half filling)
of $-1$ or $+1$.

\begin{figure}
\centering
\includegraphics[width=0.65\columnwidth]{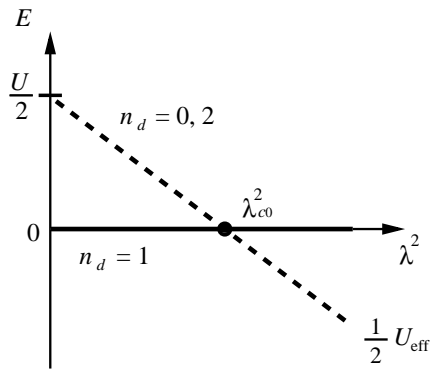}
\caption{\label{fig:imp-levels}
Symmetric, zero-hybridization model defined by $\H_{\ZH}$ in
Eq.\ \protect\eqref{H_CCBA} with $\delta_d = 0$:
evolution with \eb\ coupling $\lambda^2$ of the lowest eigenenergy
in the spin sector ($n_d=1$, solid line) and in the charge sector
($n_d=0,2$, dashed line).
A level crossing occurs at $\lambda=\lambda_{c0}$ specified in
Eq.\ \protect\eqref{lambda_c0}.}
\end{figure}

Figure \ref{fig:imp-levels} illustrates this renormalization of the Coulomb
interaction for the symmetric model ($\delta_d=0$), in which the $n_d = 0$
and $n_d = 2$ states always have the same energy. In this case, all four
impurity states become degenerate at a crossover \eb\ coupling
\begin{equation}
\label{lambda_c0}
K_0 \lambda_{c0} = \sqrt{\pi sU/2\Omega} \, .
\end{equation}
The impurity contributions to physical properties at this special point,
which is characterized by effective parameters $\Gamma=U=\epsilon_d=0$, are
identical to those at the \textit{free-orbital} fixed
point\cite{Krishna-murthy:80} of the Anderson model.

For the general case of an asymmetric impurity, the sectors $n_d = 0$ and 2
have a ground-state energy difference $E_0(n_d = 2)-E_0(n_d = 0) = 2\delta_d$
for any value of $\lambda$. The overall ground state of Eq.\ \eqref{H_CCBA}
is a doublet ($n_d=1$, $S=\pm\frac{1}{2}$) for small \eb\ couplings,
crossing over to a singlet ($n_d = 0$ for $\delta_d > 0$, or $n_d = 2$ for
$\delta_d < 0$) for large $\lambda$.
At $K_0 \lambda_{c0} = \sqrt{\pi s (U/2-|\delta_d|)/\Omega}$,
a point of three-fold ground-state degeneracy, the impurity contributions to
low-temperature ($T\ll |\delta_d|$) physical properties are identical to
those at the \textit{valence-fluctuation} fixed point\cite{Krishna-murthy:80}
of the Anderson model.

Using the NRG with only a bosonic chain [Eq.\ \eqref{H_bath:NRG}] coupled to
the impurity site, we have confirmed the existence for $\delta_d=0$ of a simple
level crossing from a spin-doublet ground state for $\lambda<\lambda_{c0}$ to
a charge-doublet ground state for $\lambda>\lambda_{c0}$. In the former regime,
the bosons couple only to the high-energy ($n_d=0$, 2) impurity states, so the
low-lying spectrum is that of free bosons obtained by diagonalizing
$H_{\bath}^{\NRG}$ given in Eq.\ \eqref{H_bath:NRG}. Here, NRG truncation plays
a negligible role provided that one works with $N_b\ge 8$ (say).

For $\lambda>\lambda_{c0}$, the low-lying bosonic excitations should, in
principle, correspond to noninteracting displaced oscillators having
precisely the same spectrum as the original bath. However, the occupation
number $a^{\dag}_{\bq} a^{\pdag}_{\bq}$ in the ground state of
Eq.\ \eqref{H_CCBA:displaced} obeys a Poisson distribution with mean
$\lambda_{\bq}^2/(N_q\,\omega_{\bq}^2)$. Thus, the total number of bosons
corresponding to operators $a_{\bq}$ satisfying
$\Omega\Lambda^{-(k+1)}<\omega_{\bq}<\Omega\Lambda^{-k}$ takes a mean value
\begin{align}
\label{n_k}
\langle \n_k\rangle_0
&= \int_{\Omega\Lambda^{-(k+1)}}^{\Omega\Lambda^{-k}} \!\! d\omega \;
   \frac{B(\omega)}{\pi\omega^2} \notag \\
&= \begin{cases}
   \displaystyle\frac{(K_0\lambda)^2}{\pi}\ln\Lambda & \text{for s = 1} \\[1ex]
   \displaystyle\frac{(K_0\lambda)^2}{\pi}
     \frac{\bigl(\Lambda^{1-s} - 1\bigr)}{(1-s)}
     \Lambda^{(1-s)k} & \text{otherwise} .
   \end{cases}
\end{align}
The bath states in the $k$th interval are represented by NRG chain states
$0\le m \le k$, with the greatest weight being borne by state $m=k$. Thus, a
faithful representation of the displaced-oscillator spectrum requires inclusion
of states having $b_m^{\dag}b_m^{\pdag}$ up to several times
$\langle\n_m\rangle_0$; based on experience with the Anderson-Holstein
model,\cite{Hewson:02+Jeon:03} one expects $N_b\ge 4\langle\n_m\rangle_0$ to
suffice. Given that $\langle\n_m\rangle_0\propto\Lambda^{(1-s)m}$, it is
feasible to meet this condition as $m\to\infty$ so long as the bath exponent
satisfies $s\ge 1$. Indeed, for Ohmic and super-Ohmic bath exponents, the NRG
spectrum for $\lambda$ not too much greater than $\lambda_{c0}$ is found to be
numerically indistinguishable from that for $\lambda = 0$. For $s<1$, by
contrast, the restriction $b_m^{\dag}b_m^{\pdag}\le N_b$ leads, for
$\lambda>\lambda_{c0}$ and large iteration numbers, to an artificially
truncated spectrum that cannot reliably access the low-energy physical
properties. Nonetheless, observation of this ``localized'' bosonic spectrum
serves as a useful indicator, both in the zero-hybridization limit and in the
full charge-coupled BFA model, that the effective \eb\ coupling remains
nonzero.

Another interpretation of Eq.\ \eqref{n_k} is that at the energy scale
$E=\Omega\Lambda^{-k}$ characteristic of interval $k$, the \eb\ coupling
takes an effective value $\tilde{\lambda}(E)$ governed by the
renormalization-group equation
\begin{equation}
\label{lambda:RG}
\frac{d\tilde{\lambda}}{d\ln(\Omega/E)} = \frac{1-s}{2} \, \tilde{\lambda},
\end{equation}
which implies that the \eb\ coupling is irrelevant for $s>1$, marginal for
$s=1$, and relevant for $s<1$. While the NRG method is capable of faithfully
reproducing the physics of $\H_{\CCBFA}$ for arbitrary renormalizations of
$\epsilon_d$, $U$, and $\Gamma$, its validity is restricted to the region
\begin{equation}
\bigl( K_0 \tilde{\lambda} \bigr)^2
  \lesssim \frac{\pi N_B}{4} \frac{1-s}{\Lambda^{1-s} - 1}
  \stackrel{\Lambda\to 1}{\longrightarrow} \frac{\pi N_B}{4\ln\Lambda} .
\end{equation}
For $\Lambda=9$ and $N_B=8$, as used in most of our calculations, the upper
limit on the ``safe'' range of $K_0\tilde{\lambda}$ varies from 1.7 for
$s=1$ to $0.9$ for $s=0$.

\begin{figure}
\centering
\includegraphics[angle=270,width=0.8\columnwidth]{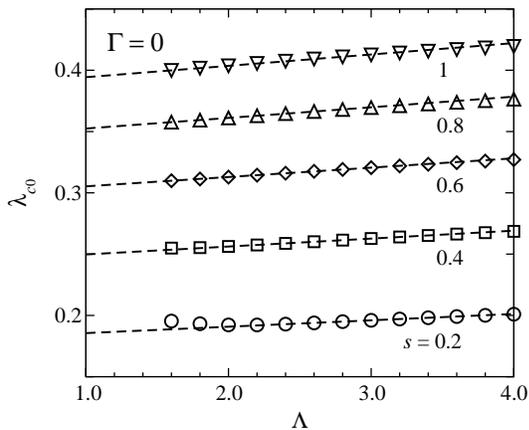}
\caption{\label{fig:lambda_c0}
Dependence of the level-crossing coupling $\lambda_{c0}$ on the discretization
$\Lambda$ for the NRG solution of $\H_{\ZH}$ [Eq.\ \protect\eqref{H_CCBA}]
with $U=0.1$, $\delta_d=0$, $N_s=200$, $N_b=16$, and five different values
of the bath exponent $s$. Dashed lines show linear fits to the data.}
\end{figure}

We now focus on the value of the crossover \eb\ coupling $\lambda_{c0}$
determined using the NRG approach.
Figure \ref{fig:lambda_c0} shows for five different bosonic bath exponents $s$
that $K_0\lambda_{c0}$ has an almost linear dependence on the NRG discretization
$\Lambda$ in the range $1.6\le\Lambda\le 4$. We believe that the rise in
$K_0\lambda_{c0}$ with $\Lambda$ reflects a reduction in the effective value of
$K_0$ arising from the NRG discretization. It is known\cite{Krishna-murthy:80}
that in NRG calculations for fermionic problems, the conduction-band density of
states at the Fermi energy takes an effective value
\begin{equation}
\label{rho_0:bar}
\rho(0) = \bar{\rho}_0 = \rho_0 / A_{\Lambda} ,
\end{equation}
where
\begin{equation}
\label{A_Lambda}
A_{\Lambda} = \frac{\ln\Lambda}{2} \, \frac{1+\Lambda^{-1}}{1-\Lambda^{-1}} \, .
\end{equation}
The general trend of the data in Fig.\ \ref{fig:lambda_c0} is consistent with
there being an analogous reduction of the bosonic bath spectral density that
requires the replacement of $K_0$ by
\begin{equation}
\bar{K}_0 = K_0/A_{\Lambda,s}
\end{equation}
when extrapolating NRG results to the continuum limit $\Lambda=1$. However,
we have not obtained a closed-form expression for $A_{\Lambda,s}$.

Table \ref{tab:lambda_c0} lists values $\lambda_{c0}(\Lambda\rightarrow 1)$
extrapolated from the data plotted in Fig.\ \ref{fig:lambda_c0}. For $s\ge 0.4$,
these values are in good agreement with Eq.\ \eqref{lambda_c0}. For $s=0.2$,
however, the extrapolated value of $\lambda_{c0}$ lies significantly above
the exact value, indicating that for given $\lambda$ the NRG underestimates
the bosonic renormalization of $U$. This is most likely another consequence of
truncating the basis on each site of the bosonic tight-binding chain.

\begin{table}
\caption{\label{tab:lambda_c0}
Crossover coupling $\lambda_{c0}$ for $\H_{\ZH}$ [Eq.\ \protect\eqref{H_CCBA}]
with $U=0.1$, $\delta_d=0$, and five different values of the bath exponent $s$:
Comparison between $\lambda_{c0}(\text{exact})$ given by
Eq.\ \protect\eqref{lambda_c0} and $\lambda_{c0}(\Lambda\!\rightarrow\! 1)$,
the extrapolation to the continuum limit of numerical values obtained for
$N_s=200$, $N_b=16$, and $1.6\le\Lambda\le 4$. Parentheses surround the
estimated nonsystematic error in the last digit.}
\begin{ruledtabular}
\begin{tabular}{llllll}
$s$ & 0.2 & 0.4 & 0.6 & 0.8 & 1.0 \\
\hline \\
$\lambda_{c0}(\text{exact})$ & 0.177 &0.251 & 0.307 & 0.355 & 0.396 \\
$\lambda_{c0}(\Lambda\!\to\! 1)$ & 0.188(4)& 0.250(2) & 0.307(2) & 0.355(2)
& 0.397(3) \\
\end{tabular}
\end{ruledtabular}
\end{table}

In analyzing our NRG results for the full charge-coupled BFA model, we attempt
to compensate for the effects of discretization and truncation by replacing
Eq.\ \eqref{Ueff0:s>0} by
\begin{equation}
\label{Ueff0:NRG}
U_{\eff}^{\NRG} = U \Bigl[ 1 - (\lambda/\lambda_{c0})^2 \Bigr] .
\end{equation}
Here, $\lambda_{c0}$ is not the theoretical value predicted in
Eq.\ \eqref{lambda_c0}, but rather is obtained from runs carried out for
$\Gamma=0$ but otherwise using the same model and NRG parameters as the
data that are being interpreted.

\subsection{Zero electron-boson coupling}
\label{subsec:lambda=0}

For $\lambda=0$, the bosonic bath decouples from the electronic degrees of
freedom, which are then described by the pure Anderson model. In this section,
we briefly review aspects of the Anderson model that will prove important in
interpreting results for the charge-coupled BFA model. For further details
concerning the Anderson model, see Refs.\ \onlinecite{Hewson:93} and
\onlinecite{Krishna-murthy:80}.

For any $\Gamma>0$, and for any $U$ and $\delta_d\equiv\epsilon_d+U/2$ (whether
positive, negative, or zero), the stable low-temperature regime of the Anderson
model lies on a line of \textit{strong-coupling} fixed points corresponding to
$\Gamma=\infty$. At any of these fixed points, the system is locked into the
ground state of the atomic Hamiltonian $\H_0$, and there are no residual degrees
of freedom on the impurity site or on site $n=0$ of the fermionic chain; the NRG
excitation spectrum is that of the Hamiltonian\cite{Krishna-murthy:80}
\begin{multline}
\label{H_SC}
\H_{\SC}^{\NRG}(V_1) = D \! \sum_{n=1}^{\infty} \sum_{\sigma}
   \tau_n \bigl( f_{n\sigma}^{\dag} f_{n-1,\sigma}^{\pdag}
   + f_{n-1,\sigma}^{\dag} f_{n\sigma}^{\pdag} \bigr) \\
   + V_1 \left( \sum_{\sigma}
      f_{1\sigma}^{\dag} f_{1\sigma}^{\pdag} - 1 \right).
\end{multline}
The coefficients $\tau_n$ are identical to those entering $\H_{\band}^{\NRG}$
[Eq.\ \eqref{H_band:NRG}], except that here $\tau_1=0$. Note that in
Eq.\ \eqref{H_SC}, the sum over $n$ begins at 1 rather than 0.

As shown in Ref.\ \onlinecite{Krishna-murthy:80}, the strong-coupling fixed
points of the Anderson model are equivalent---apart from a shift of 1 in the
ground-state charge $Q$ defined in Eq.\ \eqref{Q}---to the line of
\textit{frozen-impurity} fixed points corresponding to $\epsilon_d =\infty$,
$\Gamma=U=0$, with NRG excitation spectra described by
\begin{equation}
\label{H_FI}
\H_{\FI}^{\NRG}(V_0) = \H_{\band}^{\NRG} + V_0 \left( \sum_{\sigma}
      f_{0\sigma}^{\dag} f_{0\sigma}^{\pdag} - 1 \right).
\end{equation}
The mapping between alternative specifications of the same fixed-point
spectrum is\cite{Krishna-murthy:80}
\begin{equation}
\pi\bar{\rho}_0 V_0 = -(\pi \bar{\rho}_0 V_1)^{-1},
\end{equation}
where $\bar{\rho}_0$ [see Eq.\ \eqref{rho_0:bar}] is the effective
conduction-band density of states.

The fixed-point potential scattering is related to the ground-state impurity
charge via the Friedel sum rule,
\begin{equation}
\label{Friedel}
\langle \n_d - 1\rangle_0
  = \frac{2}{\pi} \, \text{arccot}\bigl(\pi\bar{\rho}_0 V_0\bigr)
  = \frac{2}{\pi} \, \text{arctan}\bigl(-\pi\bar{\rho}_0 V_1\bigr).
\end{equation}
For $|\delta_d|,\Gamma\ll U\ll D$, one finds that
\begin{equation}
\label{nd-1:small}
\langle \n_d - 1\rangle_0 = - \frac{8\delta_d\Gamma}{\pi A_{\Lambda} U^2},
\end{equation}
where $A_{\Lambda}$ is defined in Eq.\ \eqref{A_Lambda}.

Even though the stable fixed point of the Anderson model for any $\Gamma>0$
is one of the strong-coupling fixed points described above, the route by which
such a fixed point is reached can vary widely, depending on the relative
values of $U$, $\delta_d$, and $\Gamma$. For our immediate purposes, it
suffices to focus on the symmetric ($\delta_d = 0$) model,
for which there is a single strong-coupling fixed point corresponding
to $V_0 = \pm\infty$ or $V_1 = 0$.
If the on-site Coulomb repulsion is strong enough that the system
enters the local-moment regime ($T, \Gamma\ll U$), then it is possible to
perform a Schrieffer-Wolff transformation\cite{Schrieffer:66} that restricts
the system to the sector $n_d = 1$ and reduces the Anderson model to the Kondo
model described by the Hamiltonian
\begin{multline}
\label{H_K}
\H_{\K} = \H_{\band} + \frac{J_z}{4N_k}
  \left( \n_{d\uparrow} - \n_{d\downarrow} \right)
  \sum_{\bk,\bk'} \left( c_{\bk\uparrow}^{\dag} c_{\bk'\uparrow}^{\pdag} -
    c_{\bk\downarrow}^{\dag} c_{\bk'\downarrow}^{\pdag} \right) \\
  + \frac{J_{\perp}}{2N_k} \sum_{\bk,\bk'} \left( d_{\uparrow}^{\dag}
    d_{\downarrow}^{\pdag} c_{\bk\downarrow}^{\dag} c_{\bk'\uparrow}^{\pdag}
    + \text{H.c.} \right) ,
\end{multline}
where
\begin{equation}
\label{J_Kondo}
\rho_0 J_z = \rho_0 J_{\perp} = \frac{8\Gamma}{\pi U} .
\end{equation}
The stable fixed point is approached below an exponentially small Kondo
temperature $T_{\K}$ when the spin-flip processes associated with the $J_{\perp}$
term in $\H_{\K}$ cause the effective values of $\rho_0 J_z$ and $\rho_0 J_{\perp}$
to renormalize to strong coupling, resulting in many-body screening of the
impurity spin.

Motivated by the discussion in Sec.\ \ref{subsec:Gamma=0}, we also consider
the case of strong on-site Coulomb \textit{attraction}.
In the local-charge regime ($T, \Gamma\ll -U$), a canonical transformation
similar to the Schrieffer-Wolff transformation restricts the system to the
sectors $n_d = 0$ and $n_d = 2$, and maps the Anderson model onto a
charge Kondo model described by the Hamiltonian
\begin{multline}
\label{H_CK}
\H_{\CK} = \H_{\band} + \frac{W_d}{N_k}
  \left( \n_d - 1 \right)
  \sum_{\bk,\bk'} \left( c_{\bk\uparrow}^{\dag} c_{\bk'\uparrow}^{\pdag} +
    c_{\bk\downarrow}^{\dag} c_{\bk'\downarrow}^{\pdag} \right. \\
  \left. \rule{0ex}{2.5ex} - \delta_{\bk,\bk'} \right)
  + \frac{2 W_p}{N_k} \sum_{\bk,\bk'} \left( d_{\uparrow}^{\dag}
    d_{\downarrow}^{\dag} c_{\bk\downarrow}^{\pdag} c_{\bk'\uparrow}^{\pdag}
    + \text{H.c.} \right) ,
\end{multline}
where
\begin{equation}
\label{W_Kondo}
\rho_0 W_d = \rho_0 W_p = \frac{2\Gamma}{\pi |U|} .
\end{equation}
In this case, the stable fixed point is approached below an exponentially small
(charge) Kondo temperature $T_{\K}$ when the charge-transfer processes associated
with the $W_p$ term in $\H_{\CK}$ cause the effective values of $\rho_0 W_d$ and
$\rho_0 W_p$ to renormalize to strong coupling, resulting in many-body
screening of the impurity isospin degree of freedom [associated with the
$d$-operator terms in Eqs.\ \eqref{isospin}].

Between the opposite extremes of large positive $U$ and large negative $U$ is
a mixed-valence regime $T, |U| \ll \Gamma$ in which interactions play only a
minor role. Here, the stable fixed point is approached below a temperature of
order $\Gamma$ when the effective value of $\sqrt{\Gamma/(2\pi D)}$ scales to
strong coupling, signaling strong mixing of the impurity levels with the
single-particle states of the conduction band.

\subsection{Expectations for the full model}
\label{subsec:expectations}

Insight into the behavior of the full charge-coupled BFA model described
by Eqs.\ \eqref{H_CCBFA}--\eqref{H_impbath} can be gained by performing
a Lang-Firsov\cite{Lang:62} transformation $\H_{\CCBFA} \to \H'_{\CCBFA}
= \hat{U}^{-1} \H_{\CCBFA} \hat{U}$ with
\begin{equation}
\hat{U} = \exp\left[ (\n_d-1)
  \sum_{\bq} \frac{\lambda_{\bq}}{\sqrt{N_q}\:\omega_{\bq}}
  \bigl( a_{\bq}^{\pdag} - a_{\bq}^{\dag} \bigr) \right] .
\end{equation}
The transformation eliminates $\H_{\impbath}$, leaving
\begin{equation}
\label{H'_CCBFA}
\H'_{\CCBFA} = \H'_{\imp} + \H_{\band} + \H_{\bath}
    + \H'_{\impband},
\end{equation}
where $\H'_{\imp}$ is as defined in Eqs.\ \eqref{H'_imp} and \eqref{Ueff0},
and
\begin{multline}
\label{H'_impband}
\H'_{\impband} = \frac{1}{\sqrt{N_k}} \! \sum_{\bk,\sigma} \! \left\{ \!
   V_{\bk} \exp\!\!\left[\sum_{\bq}\frac{\lambda_{\bq}\bigl(a_{\bq}^{\pdag}
     - a_{\bq}^{\dag} \bigr)}{\sqrt{N_q}\:\omega_{\bq}} \right] \!
   c_{\bk\sigma}^{\dag} d_{\sigma}^{\pdag} \right. \\
+ \left. V_{\bk}^* \exp\!\left[- \sum_{\bq} \frac{\lambda_{\bq}
   \bigl( a_{\bq}^{\pdag} - a_{\bq}^{\dag} \bigr)}{\sqrt{N_q}\:\omega_{\bq}}
   \right] d_{\sigma}^{\dag} c_{\bk\sigma}^{\pdag} \right\} .
\end{multline}
In addition to renormalizing the impurity interaction from $U$ to $U_{\eff}$
entering $\H'_{\imp}$, the \eb\ coupling causes every hybridization event to be
accompanied by the creation and annihilation of arbitrarily large numbers of
bosons.

In the case of the Anderson-Holstein model [Eq.\ \eqref{H_AH}], various limiting
behaviors are understood.\cite{Schuttler:88} In the \textit{instantaneous limit}
$\omega_0\gg \Gamma$, the bosons adjust rapidly to any change in the impurity
occupancy; for $\lambda_0^2/\omega_0\ll U\ll \omega_0$, the physics is essentially
that of the Anderson model with $U\to U_{\eff}$, while for
$\lambda_0^2/\omega_0\gg D, U, \Gamma$, there is also a reduction from $\Gamma$
to $\Gamma\exp[-(\lambda_0/\omega_0)^2]$ in the rate of scattering between the
$n_d=0$ and $n_d=2$ sectors, reflecting the reduced overlap between the ground
states in these two sectors. In the \textit{adiabatic limit} $\omega_0\ll\Gamma$,
the phonons are unable to adjust on the typical time scale of hybridization
events, and neither $U$ nor $\Gamma$ undergoes significant renormalization.

Similar analysis for the charge-coupled BFA model is complicated by the
presence of a continuum of bosonic mode energies $\omega$, only some of which
fall in the instantaneous or adiabatic limits. Nonetheless, we can use
results for the cases $\Gamma=0$ (Sec.\ \ref{subsec:Gamma=0}) and $\lambda=0$
(Sec.\ \ref{subsec:lambda=0}), as well as those for the Anderson-Holstein
model, to identify likely behaviors of the full model. Specifically, we focus
here on the evolution with decreasing temperature of the effective Hamiltonian
describing the essential physics of the symmetric ($\epsilon_d=-U/2$) model at
the current temperature. This effective Hamiltonian is obtained under the
assumption that real excitations of energy above the ground state
$E\ge\eta T$---where $\eta$ is a number around 5, say---make a negligible
contribution to the observable properties, and thus can be integrated from the
problem.

Based on the preceding discussion, one expects that at high temperatures
$T\gg\Gamma$, the physics of the charge-coupled BFA model will be very similar
to that of the Anderson model with $U$ replaced by $\tilde{U}(\eta T)$, where
\begin{equation}
\label{Ueff}
\tilde{U}(E) = U - \frac{2}{\pi}
  \int_E^{\infty} \frac{B(\omega)}{\omega} \, d\omega .
\end{equation}
Note that $\tilde{U}(0)$ is identical to $U_{\eff}$ defined in
Eq.\ \eqref{Ueff0}. For the bath spectral density in Eq.\ \eqref{B} with $s>0$,
\begin{equation}
\label{Ueff:s>0}
\tilde{U}(E) = U - \frac{2 (K_0 \lambda)^2}{\pi s}
  \bigl[ 1-(E/\Omega)^s \bigr] \Omega .
\end{equation}
When analyzing NRG data, we instead use
\begin{equation}
\label{Ueff:NRG}
\tilde{U}^{\NRG}(E) = U \Bigl\{ 1 - (\lambda/\lambda_{c0})^2 
  \bigl[ 1-(E/\Omega)^s \bigr]\Bigr\} ,
\end{equation}
where $\lambda_{c0}$ is the empirically determined value discussed in connection
with Eq.\ \eqref{Ueff0:NRG}.

If, upon decreasing the temperature to some value $T_{\LM}$, the system comes to
satisfy $\tilde{U}(\eta T_{\LM})=\eta\max(T_{\LM},\Gamma)$, then it should
enter a local-moment regime described by the effective Hamiltonian
$\H_{\LM}=\H_{\K} + \H_{\bath}$ with the exchange couplings in $\H_{\K}$
[Eq.\ \eqref{H_K}] determined by Eq.\ \eqref{J_Kondo} with
$U\to \tilde{U}(\eta T_{\LM})$, similar to what is found in the Anderson-Holstein
model.\cite{Cornaglia:04+05} Since they couple only to the high-energy
sectors $n_d=0$ and $n_d=2$ that are projected out during the Schrieffer-Wolff
transformation, the bosons should play little further role in determining the
low-energy impurity physics.
The outcome should be a conventional Kondo effect where the \eb\ coupling
contributes only to a renormalization of the Kondo scale $T_{\K}$.

If, instead, at some $T=T_{\LC}$ the system satisfies
$\tilde{U}(\eta T_{\LC})=-\eta\max(T_{\LC},\Gamma)$, then it should enter a
local-charge regime described by the effective Hamiltonian
\begin{equation}
\label{H_LC}
\H_{\LC} = \H_{\CK} + \H_{\bath} + \H_{\impbath} .
\end{equation}
Based on the behavior of the Anderson-Holstein model,\cite{Cornaglia:04+05}
one expects $W_d$ in $\H_{\CK}$ [Eq.\ \eqref{H_CK}] to be determined by
Eq.\ \eqref{W_Kondo} with $U\to \tilde{U}(\eta T_{\LC})$,  but with $W_p$
exponentially depressed due to the aforementioned reduction in the overlap
between the ground states of the $n_d=0$ and $n_d=2$ sectors.
The bosons couple to the low-energy sector of the impurity Fock space, so they
have the potential to significantly affect the renormalization of $W_d$ and
$W_p$ upon further reduction in the temperature.
In particular, the $\lambda$ term in $\H_{\LC}$, which favors localization of
the impurity in a state of well-defined $n_d = 0$ or $2$, directly competes
with the $W_p$ double-charge transfer term that is responsible for the charge
Kondo effect of the negative-$U$ Anderson model. This nontrivial competition
gives rise to the possibility of a QPT between qualitatively distinct ground
states of the charge-coupled BFA model.

Between these extremes, the system can enter a mixed-valence regime of
small effective on-site interaction. In this regime, one must retain all
the impurity degrees of freedom of the charged-coupled BFA model. The
impurity-band hybridization competes with the \eb\ coupling for
control of the impurity, again suggesting the possibility of a QPT.

Each of the regimes discussed above features competition between
band-mediated tunneling within the manifold of impurity states and the
localizing effect of the bosonic bath. Although the tunneling is dominated
by a different process in the three regimes, it always drives the system
towards a nondegenerate impurity ground state, whereas the \eb\ coupling
favors a doubly-degenerate ($n_d = 0$, 2) impurity ground state.
In order to provide a unified picture of the three regimes (and the
regions of the parameter space that lie in between them), we will find it
useful to interpret our NRG result in terms of an overall tunneling rate
$\Delta$, which has a bare value
\begin{equation}
\label{Delta:defn}
\Delta \simeq \sqrt{ J_{\perp}^2 + 2\Gamma D/\pi + 16 W_p^2 } .
\end{equation}
Here, $W_p$ is assumed to be negligibly small in the local-moment regime,
and $J_{\perp}$ to be similarly negligible in the local-charge regime.
If $\Delta$ renormalizes to large values while the \eb\ coupling $\lambda$
scales to weak coupling, then one expects to recover the strong-coupling
physics of the Anderson model. If, on the other hand,
$\lambda$ becomes strong while $\Delta$ becomes weak, the system should enter
a low-energy regime in which the bath governs the asymptotic low-energy,
long-time impurity dynamics. Whether or not each of these scenarios is
realized in practice, and whether or not there are any other possible ground
states of the model, can be determined only by more detailed study. These
questions are answered by the NRG results reported in the sections that follow.

\section{Results: Symmetric model with sub-Ohmic dissipation}
\label{sec:sub-ohmic}

\begin{figure}
\centering
\includegraphics[width=0.65\columnwidth]{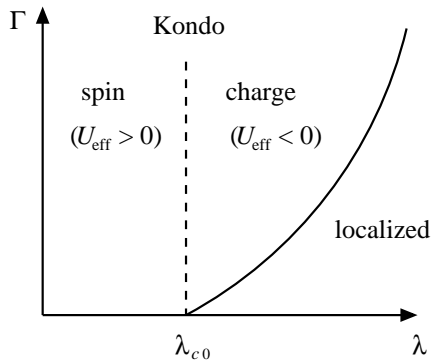}
\caption{\label{fig:phase-diagram}
Schematic phase diagram of the symmetric charge-coupled BFA model for bath
exponents $0<s<1$. The solid curve marks the boundary between the Kondo phase,
in which the impurity degrees of freedom are screened by conduction electrons,
and the localized phase, in which the impurity dynamics are controlled by the
coupling to the dissipative bath. The dashed vertical line represents a
crossover from a regime in which Kondo screening takes place primarily in the
spin sector to a regime in which a charge-Kondo effect is predominant.}
\end{figure}

This section presents results for Hamiltonian \eqref{H_CCBFA} with
$U=-2\epsilon_d > 0$ and with sub-Ohmic dissipation characterized by an
exponent $0<s<1$. Figure \ref{fig:phase-diagram} shows a schematic phase
diagram on the $\lambda$--$\Gamma$ plane at fixed $U$.
There are two stable phases: the \textit{localized} phase, in which the
impurity dynamics are controlled by the coupling to the bosonic bath and
the system has a pair of ground states related to one another by a
particle-hole transformation; and the \textit{Kondo} phase, in which there is
a nondegenerate ground state. These phases are separated by a continuous
QPT that takes place on the phase boundary (solid line in Fig.\
\ref{fig:phase-diagram}), which we parametrize as $\lambda=\lambda_c(\Gamma)$.
Within the Kondo phase, the nature of the correlations evolves continuously
with increasing $\lambda$ (at fixed $\Gamma$) from a pure spin-Kondo effect
for $\lambda = 0$ to a predominantly charge-Kondo effect beyond a crossover
(dashed line in Fig.\ \ref{fig:phase-diagram}) associated with the change in
sign of $U_{\eff}$ defined in Eq.\ \eqref{Ueff0}.

As $s$ decreases, and the \eb\ coupling becomes increasingly relevant---in a
renormalization-group sense [see Eq.\ (\ref{lambda:RG})]---the phase boundary moves to
the left as the localized phase grows at the expense of the Kondo phase, which
disappears entirely for $s\le 0$. As will be seen in Sec.\ \ref{sec:ohmic},
the phase diagram of the Ohmic ($s=1$) problem has the same topology as
Fig.\ \ref{fig:phase-diagram}, even though (as described in Sec.\ \ref{sec:ohmic})
the nature of the QPT is qualitatively different than for $0<s<1$. For $s>1$,
the \eb\ coupling is irrelevant, and the system is in the Kondo phase for all
$\Gamma>0$.

The remainder of this section presents the evidence for the previous statements.
We first discuss the renormalization-group flows and fixed points. We then turn to the behavior
in the vicinity of the phase boundary, focusing in particular on the
critical response of the impurity charge to a local electric potential.
Following that, we present results for the impurity spectral function, and
show that the low-energy scale extracted from this spectral function
supports the qualitative picture laid out in the paragraphs above and
summarized in Fig.\ \ref{fig:phase-diagram}.

\subsection{NRG flows and fixed points}
\label{subsec:NRGflows}

\begin{figure}
\centering
\includegraphics[width=0.75\columnwidth]{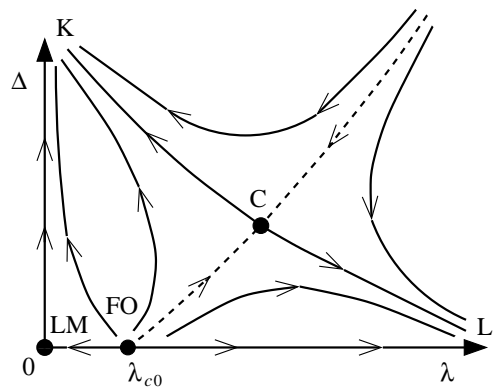}
\caption{\label{fig:RGflows}
Schematic renormalization-group flows on the $\lambda$-$\Delta$ plane for the
symmetric charge-coupled BFA model with a bath exponent $0<s<1$. Trajectories
with arrows represent the flow of the couplings $\lambda$ entering
Eq.\ \protect\eqref{H_impbath:NRG} and $\Delta$ defined in
Eq.\ \protect\eqref{Delta:defn} under decrease of the high-energy cutoffs on
the conduction band and the bosonic bath. Between the basins of attraction of
the Kondo fixed point (K) and the localized fixed point (L) lies a
separatrix, along which the flow is away from the free-orbital fixed point (FO)
located at $\lambda=\lambda_{c0}$, $\Delta=0$ and toward the critical fixed
point (C). For $\Delta=0$ only, there is flow from FO towards the local-moment
fixed point (LM) at $\lambda=0$.}
\end{figure}

Figure \ref{fig:RGflows} plots the schematic renormalization-group flows of the
couplings $\lambda$ entering Eq.\ \eqref{H_impbath:NRG} and $\Delta$ defined in
Eq.\ \eqref{Delta:defn} for a symmetric  impurity ($U=-2\epsilon_d$) coupled
to bath described by an exponent $0<s<1$.
These flows are deduced from the evolution of the many-body spectrum with
increasing iteration number $N$, i.e., with reduction in the effective band and
bath cutoffs $\tilde{D}=\tilde{\Omega}\simeq D\Lambda^{-N/2}$.
A separatrix (dashed line) forms the boundary between the basins of attraction
of a pair of stable fixed points, regions that correspond to the two phases
shown in Fig.\ \ref{fig:phase-diagram}.
Figure \ref{fig:RGflows} also shows three unstable
fixed points. In contrast to the situation at other points on the flow
diagram, each of the fixed points exhibits a many-body spectrum
that can be interpreted as the direct product of a set of bosonic
excitations and a set of fermionic excitations.

The \textit{Kondo} fixed point corresponds in the renormalization-group
language of Fig.\ \ref{fig:RGflows} to effective couplings $\lambda = 0$ and
$\Delta=\infty$. The many-body spectrum decomposes into the direct product
of (i) the excitations of a free bosonic chain described by
Eq.\ \eqref{H_bath:NRG} alone, and (ii) the strong-coupling excitations of
the Kondo (or symmetric Anderson) model, corresponding to free electrons with
a Fermi-level phase shift of $\pi/2$. This spectrum, which exhibits SU(2)
symmetry both in the spin and charge (isospin) sectors, is identical to that
found throughout the Kondo phase of the particle-hole-symmetric Ising BFK
Hamiltonian\cite{Glossop:05,Glossop:07} (a model in which the bosons couple to
the impurity's spin rather than its charge).

The schematic RG flow diagram in Fig.\ \ref{fig:RGflows} shows a
\textit{localized} fixed point corresponding to $\lambda=\infty$ and
$\Delta=0$. However, this is really a \textit{line} of fixed points described
by $\H_{\LC}$ [Eq.\ \eqref{H_LC}] with effective couplings $\lambda=\infty$,
$W_p = 0$, and $0\le W_d<\infty$. Since $W_p=0$, the impurity occupancy takes
a fixed value $n_d = 0$ or 2. (It is important to distinguish $n_d$, used to
characterize the fixed-point excitations, from the physical expectation value
of $\n_d$. The latter quantity is discussed in
Sec.\ \ref{subsec:static-response}.)

Each fixed point along the localized line has an excitation spectrum that
decomposes into the direct product of (i) bosonic excitations identical to
those at the localized fixed point of the spin-boson
model\cite{Bulla:03+05} with the same bath exponent $s$, and
(ii) fermionic excitations described by a Hamiltonian
\begin{equation}
\label{H_localized}
\H_{L,f}^{\NRG} = \H_{\band}^{\NRG} + W_d (n_d-1) \left( \sum_{\sigma}
      f_{0\sigma}^{\dag} f_{0\sigma}^{\pdag} - 1 \right),
\end{equation}
which is just the discretized version of $\H_{\CK}$ [Eq.\ \eqref{H_CK}] with
$W_p = 0$ and the operator $\n_d$ replaced by the parameter $n_d$.
The low-lying many-body eigenstates of $\H_{L,f}^{\NRG}$ appear in degenerate pairs,
one member of each pair corresponding to $n_d=0$ and the other to $n_d=2$.
The fixed-point coupling $W_d$ increases monotonically as the bare \eb\
coupling $\lambda$ decreases from infinity, and diverges on approach to
the phase boundary. As illustrated in Fig.\ \ref{fig:W_d,L}, this divergence
can be fitted to the power-law form
\begin{equation}
\label{W_d,L}
W_d \propto (\lambda-\lambda_c)^{-\beta}
  \qquad \text{for } \lambda\to\lambda_c^+.
\end{equation}
For reasons that will be explained in Sec.\ \ref{subsec:static-response},
the numerical value of $\beta$ coincides, to within a small error,
with that of the order-parameter exponent $\beta$ defined in
Eq.\ \eqref{beta:defn}.

\begin{figure}
\centering
\includegraphics[angle=270,width=0.85\columnwidth]{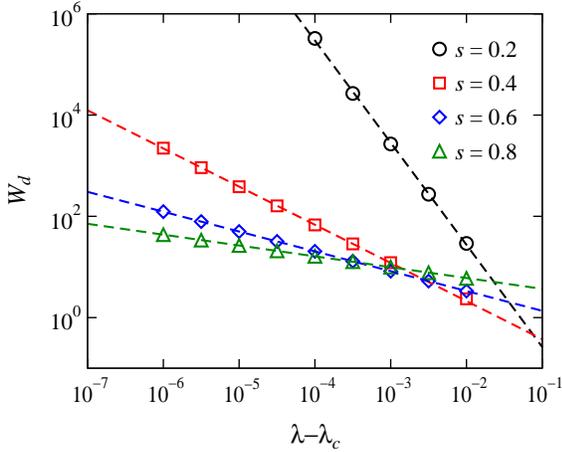}
\caption{\label{fig:W_d,L}
(Color online)
Fixed-point coupling $W_d$ entering Eq.\ \protect\eqref{H_localized} vs
\eb\ coupling $\lambda-\lambda_c$ in the localized phase near the phase
boundary at $\lambda=\lambda_c$.
Results are shown for $U=-2\epsilon_d=0.1$, $\Lambda=9$, $N_s=500$, $N_b=8$,
four different values of the bath exponent $s$, and $\Gamma=0.5$, 1.0,
10, and 50 for $s = 0.2$, 0.4, 0.6, and 0.8,
respectively (Ref.\ \protect\onlinecite{large-Gammas}). The power-law divergence of $W_d$
as $\lambda\rightarrow\lambda^{+}_c$ [Eq.\ \protect\eqref{W_d,L}] is
reflected in the linear behaviors of data on a logarithmic scale. The numerical
values of the exponent $\beta$ obtained here are identical (to within small
errors) to those listed in Table \protect\ref{tab:exponents}.}
\end{figure}

The \textit{free-orbital} fixed point ($\lambda=\lambda_{c0}$, $\Delta=0$)
is unstable with respect to a bare $\Gamma\ne 0$ or any deviation of $\lambda$
from $\lambda_{c0} \equiv \lim_{\Gamma\to 0} \lambda_c(\Gamma)$.
The \textit{local-moment} fixed point ($\lambda=\Delta=0$), at which the
impurity has a spin-$\frac{1}{2}$ degree of freedom decoupled from the band
and from the bath, is reached only for bare couplings $\Gamma=0$ (hence,
$\Delta = 0$) and $\lambda<\lambda_{c0}$.

Of greatest interest is the unstable \textit{critical} fixed point that is
reached for any bare couplings lying on the boundary
$\lambda=\lambda_c(\Gamma)$ between the Kondo and localized phases. At this
fixed point, the low-lying spectrum can be constructed as the direct product
of (i) the critical spectrum of the spin-boson model with the same bath
exponent $s$, and (ii) the strong-coupling spectrum of the Kondo (or symmetric
Anderson) model. This spectrum, which exhibits full SU(2) symmetry in both the
spin and isospin sectors, is identical to that at the critical point of the
Ising-anisotropic Bose-Fermi Kondo model,\cite{critical_decomposition} and is
illustrated in Fig.\ 3(c) of Ref.\ \onlinecite{Glossop:07}.

The decomposition of the critical spectrum can be understood by considering
the flow of couplings entering the local-charge Hamiltonian $\H_{\LC}$ defined
in Eq.\ \eqref{H_LC}. The fixed-point value of the density-density coupling is
$W_d=\infty$ in the charge-Kondo regime of the Kondo phase and diverges according
to Eq.\ \eqref{W_d,L} in the localized phase. It is therefore reasonable to assume
that in the vicinity of the phase boundary, $W_d$ rapidly renormalizes to strong
coupling, locking the impurity site and site $n=0$ of the fermionic chain into
one of just two states, which we can write in a pseudospin notation as
$|\!\!\Uparrow\rangle = d_{\uparrow}^{\dag}d_{\downarrow}^{\dag}|0\rangle$ and
$|\!\!\Downarrow\rangle = f_{0\uparrow}^{\dag}f_{0\downarrow}^{\dag}|0\rangle$,
where $|0\rangle$ is the no-particle vacuum.
Hopping of electrons on or off site $n=0$ is forbidden, so the discretized
form of $\H_{\LC}$ reduces to an effective Hamiltonian
\begin{equation}
\label{H_C}
\H_{\LC}^{\NRG}(W_d=\infty) = \H_{\SC}^{\NRG}(0) + \H_{\SBM}^{\NRG} .
\end{equation}
Here, $\H_{\SC}^{\NRG}(0)$ [Eq.\ \eqref{H_SC}] acts only on fermionic chain
sites $n\ge 1$, and yields the Kondo/Anderson strong-coupling excitation
spectrum, while
\begin{multline}
\label{H_SBM}
\H_{\SBM}^{\NRG} = \H_{\bath}^{\NRG} + 2 W_p
      \bigl( |\!\!\Uparrow\rangle\langle\Downarrow\!\!| +
             |\!\!\Downarrow\rangle\langle\Uparrow\!\!| \bigr) \\
    + \frac{\Omega K_0 \lambda}{\sqrt{\pi(s+1)}} \,
      \bigl( |\!\!\Uparrow\rangle\langle\Uparrow\!\!| -
             |\!\!\Downarrow\rangle\langle\Downarrow\!\!| \bigr) \,
      \bigl( b_0^{\pdag} + b_0^{\dag} \bigr)
\end{multline}
acts on the remaining degrees of freedom in the problem in a subspace of
states all carrying quantum numbers $S=S_z=Q=0$.
$\H_{\SBM}^{\NRG}$ is precisely the discretized form of the spin-boson
Hamiltonian with tunneling rate $\Delta=4 W_p$ and dissipation strength
$\alpha=2(K_0 \lambda)^2/\pi$.
These two couplings compete with one another, with three possible
outcomes: (1) $\Delta$ can scale to infinity and $\alpha$ to zero,
resulting in flow to the delocalized fixed point (the Kondo fixed point of the
charge-coupled BFA model); (2) $\alpha$ can scale to infinity and $\Delta$
to zero, yielding flow to the localized fixed point;
or (3) both couplings can renormalize to finite values $\Delta=\Delta_C$,
$\alpha=\alpha_C$ at the critical point. This picture implies that the
universal critical behavior of the charge-coupled BFA model should be
identical to that of the spin-boson model, the conduction-band electrons
serving only to dress the $n_d=0, \, 2$ impurity levels and to renormalize
the impurity tunneling rate and the dissipation strength.

\begin{figure}
\centering
\includegraphics[angle=270,width=0.95\columnwidth]{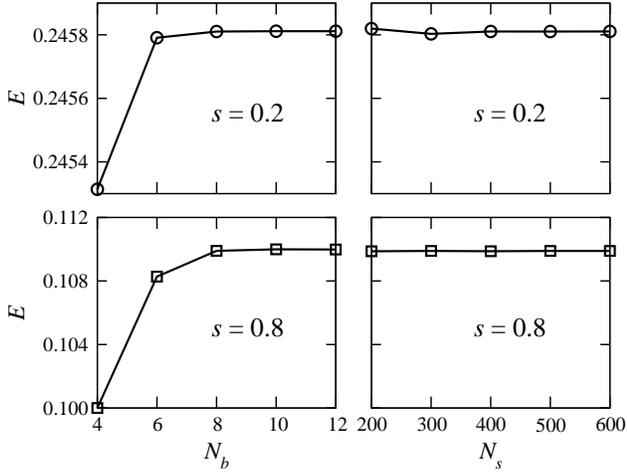}
\caption{\label{fig:E-crit}
Dependence of the energy of the first bosonic excitation at the critical point
($\lambda=\lambda_c$) on the NRG truncation parameters $N_b$ and $N_s$.
Results are shown for $U=-2\epsilon_d=0.1$, $\Gamma=0.01$, $\Lambda=9$, and
bath exponents $s=0.2$ and $s=0.8$. In the left panels, $N_s=500$, while in the
right panels $N_b = 8$.}
\end{figure}

Given that the NRG approach necessarily involves Fock-space truncation, it is
instructive to examine the dependence of the fixed-point spectra on the
parameters $N_s$ and $N_b$ denoting, respectively, the number of states retained
from one NRG iteration to the next and the maximum number of bosons allowed per
site of the bosonic chain.
Figure~\ref{fig:E-crit} shows, for representative bath exponents $s=0.2$ and
$s=0.8$, that the energy of the lowest bosonic excitation at $\lambda=\lambda_c$
converges rapidly with increasing $N_s$ and $N_b$. This behavior
suggests that for $\Lambda=9$, at least, $N_s=500$ and $N_b=8$ are sufficient
for studying the physics at the critical point.

\begin{figure}
\centering
\includegraphics[angle=270,width=0.95\columnwidth]{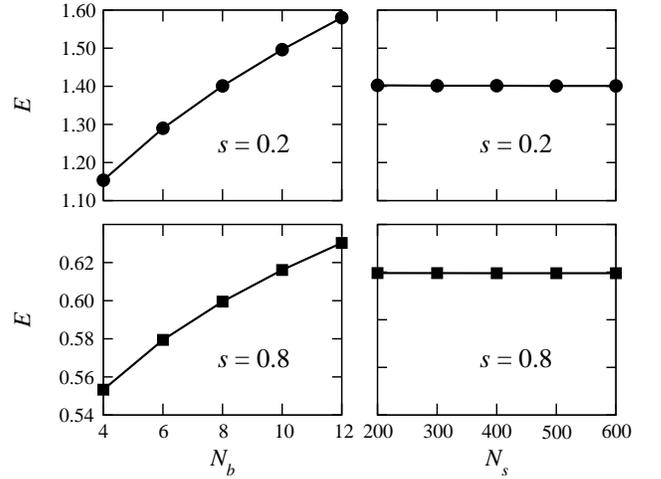}
\caption{\label{fig:E-loc}
Dependence of the energy of the first bosonic excitation in the localized phase
($\lambda=1.1\lambda_c$) on the NRG truncation parameters $N_b$ and $N_s$.
All other parameters are as in Fig.\ \protect\ref{fig:E-crit}.}
\end{figure}

By contrast, the lowest bosonic excitation energy for $\lambda=1.1\lambda_c$,
plotted in Fig.\ \ref{fig:E-loc}, converges only slowly with respect to $N_b$.
This points to the failure of the truncated bosonic basis deep inside the
localized phase of the sub-Ohmic model, where the mean boson number per site is
expected to diverge according to Eq.\ \eqref{n_k}. This interpretation is
confirmed by calculation of the expectation value of the total boson number,
\begin{equation}
\hat{B}_N=\sum_m^{M(N)} b_m^{\dag} b_m^{\pdag},
\end{equation}
where $M(N)$ denotes the highest labeled bosonic site present at iteration $N$.
Our results for $\langle\hat{B}_{20}\rangle$ vs $N_b$ (not shown) are very
similar to those in Fig.\ 8 of Ref.\ \onlinecite{Glossop:07}, with
convergence by $N_b = 8$ at the critical point, but no evidence of such
convergence for an \eb\ coupling 10\% over the critical value.

Recently, Bulla et al.\ applied a ``star'' reformulation of the NRG
to the spin-boson model.\cite{Bulla:03+05} While this approach provides a
good description of the localized fixed point, it does not correctly capture
the physics of the delocalized phase (corresponding to the Kondo phase of the
present model) or of the critical point that separates the two stable phases.
For this reason, we prefer to work with the ``chain'' formulation summarized
in Sec.\ \ref{sec:model}.

\subsection{Critical coupling}
\label{subsec:critical-coupling}

\begin{figure}
\centering
\includegraphics[angle=270,width=0.8\columnwidth]{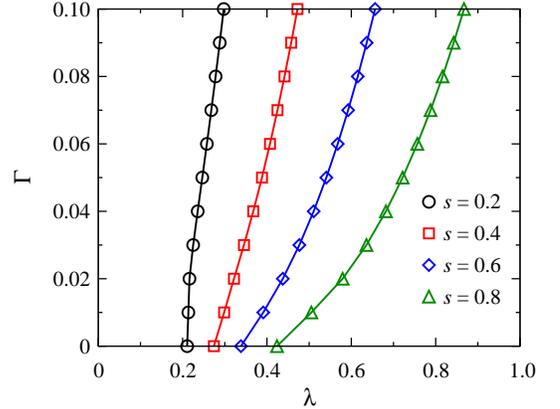}
\caption{\label{fig:lambda_c-vs-Gamma}
(Color online)
Critical coupling $\lambda_c$ vs hybridization width $\Gamma$ for
$U=-2\epsilon_d=0.1$, $\Lambda=9$, $N_s=500$, $N_b=8$, and the bath
exponents $s$ listed in the legend.}
\end{figure}

Figure~\ref{fig:lambda_c-vs-Gamma} plots the critical \eb\ coupling
$\lambda_c(\Gamma)$ for fixed $U=-2\epsilon_d$ and four different
values of the bath exponent $s$.
As expected, with increasing $\Gamma$, the critical coupling increases
smoothly from $\lambda_c(\Gamma=0) \equiv \lambda_{c0}$, reflecting the
fact that entry to the localized phase requires an \eb\ coupling
sufficiently large not only to drive $U_{\eff}$ negative, but also to
overcome the reduction in the electronic energy that derives from the
hybridization. We believe that the vertical slope of the $s=0.2$
phase boundary as it approaches the horizontal axis in
Fig.~\ref{fig:lambda_c-vs-Gamma} is an artifact stemming from the
same source as the NRG overestimate of $\lambda_{c0}$ for the same
bath exponent. (See the discussion of Fig.\ \ref{fig:lambda_c0} in Sec.\
\ref{subsec:Gamma=0}.)

In the subsections that follow, we show that the critical properties of the
charge-coupled BFA model map, under interchange of spin and charge degrees
of freedom, onto those of the spin-coupled BFA model studied (along with the
corresponding Ising BFK model) in Ref.\ \onlinecite{Glossop:07}. The
spin-coupled model is described by Eqs.\ \eqref{H_CCBFA}--\eqref{H_impband}
and \eqref{H_band:NRG}--\eqref{H_impband:NRG}, with Eqs.\ \eqref{H_impbath}
and \eqref{H_impbath:NRG} replaced by
\begin{equation}
\label{H_impbath:spin-BFAM}
\H_{\impbath}
  = \frac{1}{2\sqrt{N_q}}(\n_{d\uparrow} - \n_{d\downarrow})
    \sum_{\bq} g_{\bq} \ \bigl( a_{\bq}^{\pdag} + a_{-\bq}^{\dag} \bigr)
\end{equation}
and
\begin{equation}
\label{H_impbath:spin-BFAM:NRG}
\H_{\impbath}^{\NRG}
  = \frac{\Omega K_0 g}{2\sqrt{\pi(s+1)}} \,
    (\n_{d\uparrow} - \n_{d\downarrow})
    \bigl( b_0^{\pdag} + b_0^{\dag} \bigr) .
\end{equation}
In light of the parallels between the universal critical behavior of the two
models, it is of interest to compare their critical couplings, making due
allowance for the additional prefactor of $\frac{1}{2}$ that enters
Eqs.\ \eqref{H_impbath:spin-BFAM} and \eqref{H_impbath:spin-BFAM:NRG}.

\begin{figure}
\centering
\includegraphics[angle=270,width=0.8\columnwidth]{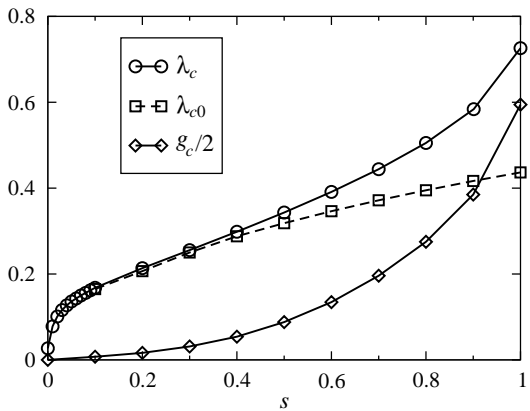}
\caption{\label{fig:lambda_c-vs-s}
Variation with bath exponent $s$ of the critical couplings $\lambda_c$ and
$\lambda_{c0}$ in the charge-coupled BFA model (this work) and $g_c/2$ in the
spin-coupled BFA model (Ref.\ \onlinecite{Glossop:07}). Results are shown
for $U=-2\epsilon_d=0.1$, $\Gamma=0.01$, $\Lambda=9$, $N_s=500$, and $N_b=8$.}
\end{figure}

Figure~\ref{fig:lambda_c-vs-s} plots the $s$ dependence of
$\lambda_c$ and $g_c/2$ for fixed values of $U = -2\epsilon_d$ and $\Gamma$.
For all $0<s\le 1$, $\lambda_c$ is found to exceed $g_c/2$. This fact
can be understood by noting the contrasting role of the \eb\ coupling in
the two models.
In the spin-coupled BFA model, increasing $g$ from zero immediately begins to
localize the impurity in a state of fixed $S_z$, and thereby to impede the
spin-flip processes that are central to the Kondo effect.
In the charge-coupled model, by contrast, increasing $\lambda$ from zero
initially acts to decrease the effective Coulomb repulsion and hence to enhance
charge fluctuations on the impurity site; only for
$\lambda_c \gtrsim \lambda_{c0}$ do further increases in the \eb\ coupling
serve to localize the impurity in a state of fixed charge, eventually leading
to the suppression of the charge Kondo effect at $\lambda=\lambda_c$.

\subsection{Crossover scale}
\label{subsec:T*}

Under the renormalization-group flows sketched in Fig.\ \ref{fig:RGflows},
the system passes, with decreasing energy cutoff or decreasing temperature,
between the regions of influence of different renormalization-group fixed
points. For bare parameters that place the system near the boundary between
the Kondo and localized phases, the free-orbital fixed point typically governs
the behavior at temperatures much greater than the Kondo temperature $T_{\K}$ of
the Anderson model obtained by setting $\lambda=0$ in Eq.\ \eqref{H_CCBFA}.
For temperatures between of order $T_{\K}$ and a crossover scale $T_*$, the
system exhibits quantum critical behavior controlled by thermal fluctuations
about the unstable critical point. Finally, the physics in the regime
$T\lesssim T_*$ is governed by one or other of the two stable fixed points:
Kondo or localized.

For fixed values of all other parameters, one expects $T_*$ to vanish as the
\eb\ coupling approaches its critical value according to a power law:
\begin{equation}
\label{nu:defn}
T_* \propto |\lambda-\lambda_c|^{\nu}
  \qquad \text{for } \lambda \to \lambda_c ,
\end{equation}
where $\nu$ is the correlation-length exponent.\cite{Sachdev:99}
The crossover scale can be determined directly from the NRG solution via the
condition $T_*\propto\Lambda^{-N_*/2}$, where $N_*$ is the number of the
iteration at which the many-body energy levels cross over to those of a stable
fixed point. There is some arbitrariness as to what precisely constitutes
crossover of the levels. Different criteria will produce $T_*(\lambda)$ values
that differ from one another by a $\lambda$-independent multiplicative factor.
It is of little importance what definition of $N_*$ one uses, provided that
it is applied consistently.

\begin{figure}
\centering
\includegraphics[angle=270,width=0.85\columnwidth]{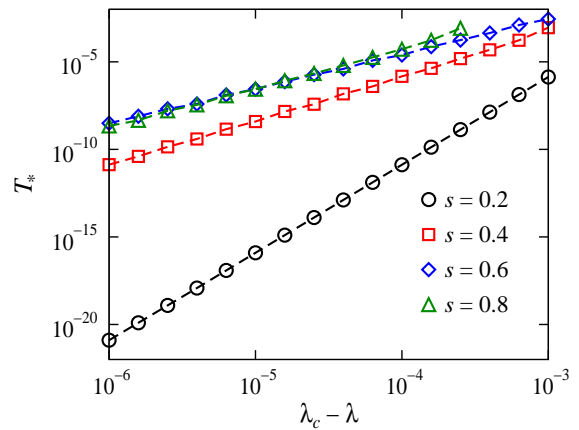}
\caption{\label{fig:nu}
(Color online)
Crossover scale $T_*$ vs $\lambda_c-\lambda$ on the Kondo side of the
critical point for four different values of the bath exponent $s$, with all
other parameters as in Fig.\ \protect\ref{fig:W_d,L}. The slope of each line
on this log-log plot gives the correlation-length exponent $\nu(s)$ defined in
Eq.\ \protect\eqref{nu:defn}.}
\end{figure}

Figure \ref{fig:nu} shows typical dependences of $T_*$ on
\mbox{$\lambda_c-\lambda$}
in the Kondo phase. Equation \eqref{nu:defn} holds very well over several
decades, as demonstrated by the linear behavior of the data on a log-log plot.
We find that the numerical values of $\nu(s)$, some of which are listed in
Table \ref{tab:nu}, are identical (within small errors), to those of the
spin-boson and Ising BFK models for the same bath exponent $s$. This supports
the notion that the critical point of the charge-coupled BFA model belongs to
the same universality class as the critical points of the spin-boson and Ising
BFK models. However, to confirm this equivalence, we must compare other
critical exponents, as reported below.

\begin{table}
\caption{\label{tab:nu}
Correlation-length critical exponent $\nu$ vs bath exponent $s$ for the
charge-coupled Bose-Fermi Anderson model (CC-BFA, this work) and for the
Ising-anisotropic Bose-Fermi Kondo model (BFK, from
Refs.\ \protect\onlinecite{Glossop:05} and \protect\onlinecite{Glossop:07}).
Parentheses surround the estimated nonsystematic error in the last digit.}
\begin{ruledtabular}
\begin{tabular}{c@{\hspace{3ex}}|cccc}
$s$           & 0.2     & 0.4     & 0.6     & 0.8 \\
$\nu$(CC-BFA) & 4.99(3) & 2.52(2) & 1.97(4) & 2.12(6) \\
$\nu$(BFK)   & 4.99(5) & 2.50(1) & 1.98(3) & 2.11(2)
\end{tabular}
\end{ruledtabular}
\end{table}

\subsection{Thermodynamic susceptibilities}
\label{subsec:thermodynamics}

In this subsection, we consider the response of the charge-coupled BFA model
to a global magnetic field $H$ and to a global electric potential $\Phi$.
These external probes enter the Hamiltonian through an additional term
\begin{equation}
\H_{\text{ext}} = H S_z + \Phi Q ,
\end{equation}
where $S_z$ and $Q$ are defined in Eqs.\ \eqref{S_z} and \eqref{Q},
respectively. In particular, we focus on the static impurity spin susceptibility
$\chi_{s,\imp} = -\partial^2 F_{\imp}/\partial H^2$ and the static impurity
charge susceptibility $\chi_{c,\imp} = \partial^2 F_{\imp}/\partial \Phi^2$.
Here, $F_{\imp}=\Delta(F)$, where $\Delta(X)$ is the difference between (i) the
value of the bulk property $X$ when the impurity is present and (ii) the value
of $X$ when the impurity is removed from the system. It is straightforward to
show that
\begin{align}
T\chi_{s,\imp}
&= \Delta \bigl(\langle\!\langle\hat{S}_z^2\rangle\!\rangle -
\langle\!\langle \hat{S}_z\rangle\!\rangle^2\bigr) , \\
T\chi_{c,\imp}
&= \Delta \bigl( \langle\!\langle\hat{Q}^2\rangle\!\rangle -
   \langle\!\langle\hat{Q}\rangle\!\rangle^2 \bigr) ,
\end{align}
where, for any operator $\hat{A}$,
\begin{equation}
\langle\!\langle\hat{A}\rangle\!\rangle
= \frac{\text{Tr}\,\hat{A}\exp(-\H/T)}{\text{Tr}\,\exp(-\H/T)} \, .
\end{equation}
Note that with the above definitions,
$\lim_{T\to\infty}T\chi_{s,\imp}=\frac{1}{8}$ but
$\lim_{T\to\infty}T\chi_{c,\imp}=\frac{1}{2}$, a factor of four
difference that must be taken into account when comparing the two
susceptibilities. Since each $T\chi_{\imp}$
is calculated as the difference of bulk quantities, its evaluation using
the NRG method is complicated by significant discretization and truncation
errors. In order to obtain reasonably well-converged results for $T\chi_{\imp}$,
we retain $N_s=2000$ states after each NRG iteration. However, even this
number is insufficient to allow reliable extraction of $\chi_{\imp}\equiv
(T\chi_{\imp}) / T$ as $T\to 0$.

\begin{figure}
\centering
\includegraphics[angle=270,width=0.95\columnwidth]{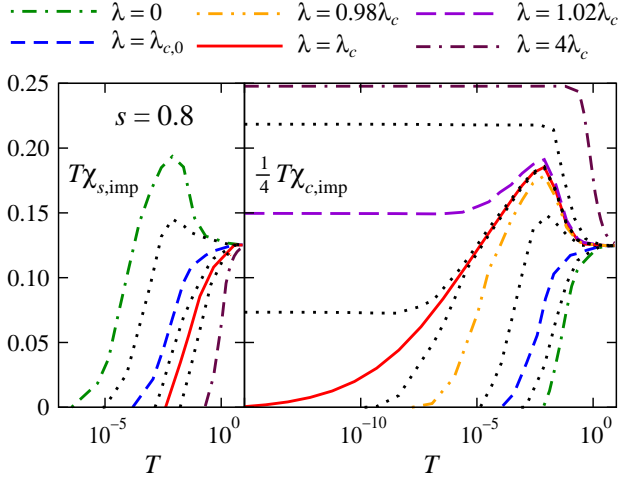}
\caption{\label{fig:Tchi-vs-T}
(Color online)
Temperature dependence of the impurity contribution to the static spin (left)
and charge (right) susceptibilities for $s=0.8$, $U=-2\epsilon_d=0.1$,
$\Gamma=0.01$, $\Lambda=9$, $N_s=2000$, $N_b=8$, and different values of
the \eb\ coupling $\lambda$.
Dotted curves correspond to \eb\ couplings lying between the $\lambda$
values specified in the legend for the adjacent nondotted curves.
For $\lambda=\lambda_{c0}\simeq 0.396$, the spin and charge susceptibilities
are equivalent: $\chi_{s,\imp}(T) \simeq \frac{1}{4}\chi_{c,\imp}(T)$.
For $\lambda<\lambda_{c0}$, the spin response is stronger,
while for $\lambda>\lambda_{c0}$, the charge response dominates.
For $\lambda\le\lambda_c\simeq 0.5052181$, $\lim_{T\to 0} T\chi_{c,\imp}(T)=0$,
whereas for $\lambda>\lambda_c$, the limiting value is nonzero and obeys
Eqs.\ \protect\eqref{delta_0-vs-W_d} and \protect\eqref{Tchi_c-vs-delta_0}.}
\end{figure}

Figure \ref{fig:Tchi-vs-T} plots NRG results for $T\chi_{s,\imp}(T)$ and
$\frac{1}{4}T\chi_{c,\imp}(T)$, calculated for bath exponent $s=0.8$ and
different values of the \eb\ coupling $\lambda$.
For $\lambda\ll\lambda_{c0}$ (see Sec.\ \ref{subsec:Gamma=0}), both
impurity susceptibilities behave very much as
they do in the Anderson model: with decreasing temperature, $T\chi_{c,\imp}$
quickly falls toward zero, signaling quenching of charge fluctuations upon
entry into the local-moment regime, whereas $T\chi_{s,\imp}$ initially rises
towards its local-moment value of $\frac{1}{4}$, before dropping to zero for $T\ll T_*$
on approach to the Kondo fixed point. With increasing $\lambda$, the charge
response grows and the spin response is suppressed. The two susceptibilities
are approximately equivalent for $\lambda = \lambda_{c0}$, where the effective
Coulomb interaction $U_{\eff}=0$. For still stronger \eb\ couplings,
$T\chi_{s,\imp}$ plunges rapidly as the temperature is decreased, whereas
$T\chi_{c,\imp}$ first rises on entry to the local-charge regime before
dropping to satisfy
\begin{equation}
\label{chi_c,imp:K}
\lim_{T\to 0} T\chi_{c,\imp}(T) = 0
\quad \text{for } \lambda < \lambda_c .
\end{equation}
These trends are very similar to those exhibited\cite{Zitko:06} by the
Anderson-Holstein model. In that model, however, the drop in
$T\chi_{c,\imp}(T)$ takes place\cite{Cornaglia:04+05} for strong \eb\
couplings $\lambda_0 \gg \sqrt{\omega_0 U/2}$ around an effective
Kondo temperature $T_K^{\eff}\sim D\exp(-\pi \lambda_0^4 /\Gamma \omega_0^3)$.
In the charge-coupled BFA model, by contrast, neither the spin susceptibility
nor the charge susceptibility exhibits any obvious feature that correlates
with the vanishing of $T_*$ as $\lambda\to\lambda_c^-$.
This can be understood by noting that the impurity susceptibilities are
determined purely by the fermionic part of the excitation spectrum, whose
asymptotic low-energy form is the same at the critical fixed point (which
governs the behavior in the quantum critical regime $T_*\lesssim T\lesssim T_K$)
as at the Kondo fixed point (which controls the regime $T\lesssim T_*$).

The behavior of the static impurity spin susceptibility is qualitatively
unchanged upon crossing from the Kondo phase to the localized phase. However,
for $\lambda>\lambda_c$, $T\chi_{c,\imp}$ approaches at low temperatures a
nonzero value that can be inferred from the effective Hamiltonian
$\H_{L,f}^{\NRG}$ [Eq.\ \eqref{H_localized}]. Electrons near the Fermi level
experience an $s$-wave phase shift
\begin{equation}
\label{delta(0)}
\delta(\omega=0) = \begin{cases}
   \delta_0 & \text{for } n_d = 0 \\
   \pi - \delta_0 & \text{for } n_d = 2 ,
   \end{cases}
\end{equation}
where $n_d$ labels the two disconnected sectors of $\H_{L,f}^{\NRG}$, and
\begin{equation}
\label{delta_0-vs-W_d}
\delta_0 = \text{arctan} \, (\pi \bar{\rho}_0 W_d), \quad
  0\le\delta_0\le\pi/2,
\end{equation}
with $\bar{\rho}_0$ being the effective conduction-band density of states
defined in Eq.\ \eqref{rho_0:bar}. It is then straightforward to show
that
\begin{equation}
\label{Tchi_c-vs-delta_0}
\lim_{T\to 0} T\chi_{c,\imp}(T) = (1 - 2 \delta_0 / \pi )^2 .
\end{equation}
Equations \eqref{W_d,L}, \eqref{delta_0-vs-W_d}, and \eqref{Tchi_c-vs-delta_0}
together imply that
\begin{equation}
\label{chi_c,imp:L}
\lim_{T\to 0} T\chi_{c,\imp}(T) \propto (\lambda - \lambda_c)^{2\beta}
\quad \text{for } \lambda \to \lambda_c^+ .
\end{equation}

As this example illustrates, the thermodynamic susceptibilities contain
signatures of an evolution from a spin-Kondo effect to a charge-Kondo effect.
Furthermore, Eqs.\ \eqref{chi_c,imp:K} and \eqref{chi_c,imp:L} suggest
that $\chi_{c,\imp}$ may serve as the order-parameter susceptibility for the
QPT. However, neither susceptibility manifests the vanishing of the crossover
scale $T_*$ on approach to the transition from the Kondo side. Moreover, the
conservation of $Q$ prevents $\chi_{c,\imp}$ from acquiring an anomalous
temperature dependence in the quantum-critical regime.\cite{Sachdev:94} Thus,
one is led to conclude that the response to a global electric potential $\Phi$
does not provide access to the critical fluctuations near the QPT.

\subsection{Local charge response}
\label{subsec:local-response}

Given the nature of the coupling in Hamiltonian \eqref{H_CCBFA} between
the impurity and the bosonic bath, we expect to be able to probe the quantum
critical point through the system's response to a local electric potential
$\phi$ that acts solely on the impurity charge, entering the Hamiltonian
via an additional term
\begin{equation}
\label{H_c,loc}
\H_{c,\loc} = \phi \, (\n_d - 1).
\end{equation}
A nonzero $\phi$ is equivalent to a shift in $\delta_d$ entering
Eq.\ \eqref{H_imp:delta} away from its bare value $\epsilon_d+U/2=0$.

In this subsection we show that for sub-Ohmic bath exponents $0<s<1$,
(i) the response to a static $\phi$ is described by critical exponents that
satisfy hyperscaling relations characteristic of an interacting quantum
critical point,
(ii) numerical values of these critical exponents are identical to those of the
spin-boson and Ising BFK models, and
(iii) the dynamical response is consistent with the presence of $\omega/T$
scaling in the vicinity of the quantum critical point.

\subsubsection{Static local charge response}
\label{subsec:static-response}

The response to imposition of a static local potential $\phi$ is measured by
the thermodynamic average value of the impurity charge,
\begin{equation}
Q_{\loc}=\langle\!\langle \n_d-1\rangle\!\rangle,
\end{equation}
and through the static local charge susceptibility
\begin{equation}
\label{static-chi}
\chi_{c,\loc}(T;\omega=0)
  = -\left.\frac{\partial Q_{\loc}}{\partial\phi}\right|_{\phi=0}
  =-\lim_{\phi\to 0}\frac{Q_{\loc}}{\phi}.
\end{equation}
In NRG calculations of $\lim_{\phi\to 0} Q_{\loc}(\phi)$ and
$\chi_{c,\loc}$, we use potentials in the range
$10^{-13}\leq |\phi| \leq 10^{-10}$.

\begin{figure}
\centering
\includegraphics[angle=270,width=0.85\columnwidth]{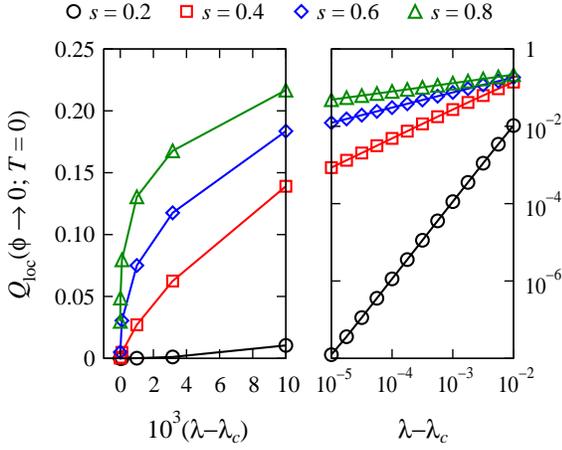}
\caption{\label{fig:beta}
(Color online)
Impurity charge $\lim_{\phi\to 0^-} Q_{\loc}(\lambda,\phi;T=0)$ vs \eb\
coupling $\lambda-\lambda_c$ for four different values of the bath exponent
$s$. All other parameters are as in Fig.\ \protect\ref{fig:W_d,L}.
As $\lambda$ approaches $\lambda_c$ from above,
$\lim_{\phi\to 0^-} Q_{\loc}(\lambda,\phi;T=0)$ vanishes (left panel) in a
power-law fashion (right panel) described by Eq.\ \protect\eqref{beta:defn}.}
\end{figure}

As illustrated in Fig.\ \ref{fig:beta}, the ``spontaneous impurity
charge'' $\lim_{\phi\to 0} Q_{\loc}(\lambda,\phi;T=0)$ indeed serves as
an order parameter for the QPT between the Kondo and
localized phases. This quantity vanishes for all $\lambda<\lambda_c$
and is nonzero for $\lambda>\lambda_c$, its onset being described
by the power law
\begin{equation}
\label{beta:defn}
\lim_{\phi\to 0} Q_{\loc}(\lambda,\phi;T=0) \propto
  (\lambda-\lambda_{c})^{\beta}
  \quad \text{for } \lambda \to \lambda_c^+ .
\end{equation}
In the localized phase, the presence of an infinitesimal local potential
restricts the effective Hamiltonian \eqref{H_localized} to just one $n_d$
sector: $n_d = 0$ for $\phi>0$, or $n_d = 2$ for $\phi<0$.
Then substituting Eq.\ \eqref{delta(0)} into the Friedel sum rule
$\langle \n_d \rangle_0 = 2\delta(0)/\pi$ yields
\begin{equation}
\label{Q_loc-vs-W_d}
\lim_{\phi\to 0} Q_{\loc}(\phi; T=0) = -\frac{2\,\sgn\phi}{\pi} \,
  \text{acot}\bigl(\pi\bar{\rho}_0 W_d\bigr) .
\end{equation}
The latter relation explains the equality of the exponents
$\beta$ entering Eqs.\ \eqref{W_d,L} and \eqref{beta:defn}.
It should also be noted that Eqs.\ \eqref{delta_0-vs-W_d},
\eqref{Tchi_c-vs-delta_0}, and \eqref{Q_loc-vs-W_d} together imply that
\begin{equation}
\lim_{\phi\to 0} Q^2_{\loc}(\phi; T=0) =
\lim_{T\to 0} T\chi_{c,\imp}(T) .
\end{equation}

\begin{figure}
\centering
\includegraphics[angle=270,width=0.8\columnwidth]{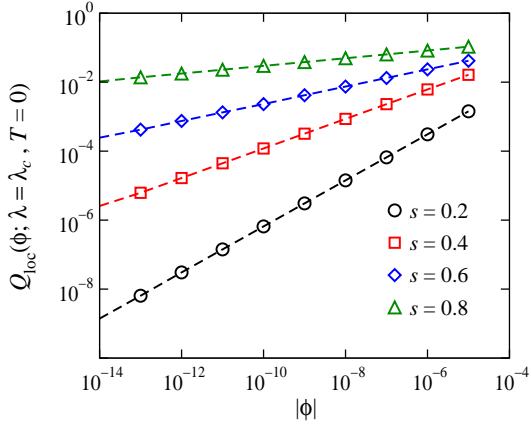}
\caption{\label{fig:delta}
(Color online)
Impurity charge $Q_{\loc}(\phi;\lambda\!=\!\lambda_c,T\!=\!0)$ vs local
electric potential $|\phi|$ for four different values of the bath exponent $s$.
All other parameters are as in Fig.\ \protect\ref{fig:W_d,L}. The dashed
lines represent fits to the form of Eq.\ \protect\eqref{delta:defn}.}
\end{figure}

At the critical point, the response to a small-but-finite potential $\phi$
obeys another power law,
\begin{equation}
\label{delta:defn}
Q_{\loc}(\phi;\lambda=\lambda_{c},T=0)
  \propto |\phi|^{1/\delta} .
\end{equation}
This behavior is exemplified in Fig.\ \ref{fig:delta} for four different
values of $s$.

\begin{figure}
\centering
\includegraphics[angle=270,width=0.85\columnwidth]{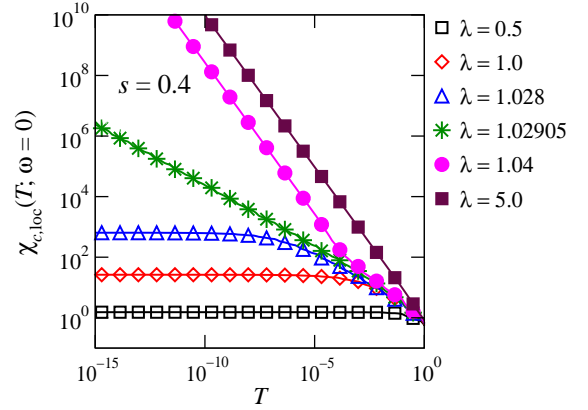}
\caption{\label{fig:static-chi}
(Color online)
Static local charge susceptibility $\chi_{c,\loc}(T;\omega=0)$ vs
temperature $T$ for $s=0.4$, $U=-2\epsilon_d=0.1$, $\Gamma=1.0$ (see footnote
\protect\onlinecite{large-Gammas}), $\Lambda=9$, $N_s=500$, $N_b=8$, and for
different values of the \eb\ coupling $\lambda$ straddling the critical
value $\lambda_c \simeq 1.02905$.}
\end{figure}

Figure \ref{fig:static-chi} shows a logarithmic plot of the static local charge
susceptibility $\chi_{c,\loc}(T;\omega=0)$ vs temperature $T$ for bath exponent
$s=0.4$ and a number of \eb\ couplings straddling $\lambda_c$.
In the quantum-critical regime, the susceptibility
has the anomalous temperature dependence
\begin{equation}
\label{x:defn}
\chi_{c,\loc}(T;\omega=0)\propto T^{-x} \quad
  \text{for } T_*\ll T\ll T_{\K},
\end{equation}
characterized by a critical exponent $x$.
For $T\ll T_*(\lambda)$, the temperature variation approaches that of one or other
of the stable fixed points. In the Kondo phase, the susceptibility is essentially
temperature independent, signaling complete quenching of the impurity, and the
zero-temperature value diverges on approach to the critical coupling as
\begin{equation}
\label{gamma:defn}
\chi_{c,\loc}(\lambda;\omega=T=0) \propto(\lambda_{c}-\lambda)^{-\gamma} \quad
  \text{for } \lambda \to \lambda_c^-.
\end{equation}
In the localized phase, by contrast,
\begin{multline}
\label{Curie-law}
\chi_{c,\loc}(T,\lambda;\omega=0) =
  \lim_{\phi\to 0} \frac{Q_{\loc}^2(\lambda,\phi;T=0)}{T} \\
  \text{for } \lambda > \lambda_c \text{ and } T\ll T_* ,
\end{multline}
indicative of a residual impurity degree of freedom.
Precisely at the critical \eb\ coupling, Eq.\ \eqref{x:defn} is obeyed all
the way down to $T=0$.

Table~\ref{tab:exponents} lists the numerical values of the critical exponents
$\beta$, $1/\delta$, $x$, and $\gamma$, for four different sub-Ohmic bath
exponents $s$. For each $s$, these critical exponents are identical within
estimated error to those of the spin-boson and Ising BFK models. In all cases,
we find that $x=s$ to within our estimated nonsystematic numerical
error. We also note that for $s\le \frac{1}{2}$, the value of $\gamma$ lies
close to its mean-field value of 1. It is conceivable that the deviations
of $\gamma$ from 1 are artifacts of the NRG discretization and truncation
approximations.

The exponents in Table~\ref{tab:exponents} obey the hyperscaling relations
\begin{equation}
\delta=\frac{1+x}{1-x} , \quad 2\beta=\nu(1-x) , \quad \gamma=\nu x ,
\end{equation}
which are consistent with the ansatz
\begin{equation}
\label{ansatz}
F=T f \left( \frac{|\lambda-\lambda_{c}|}{T^{1/\nu}} \, ,
             \frac{|\phi|}{T^{(1+x)/2}} \right)
\end{equation}
for the nonanalytic part of the free energy.
Such hyperscaling suggests that the quantum critical point is an interacting
one.\cite{Sachdev:99}

\begin{table}
\caption{\label{tab:exponents}
Static critical exponents $\beta$, $1/\delta$, $x$, and $\gamma$ defined in
Eqs.\ \protect\eqref{beta:defn} and
\protect\eqref{delta:defn}--\protect\eqref{gamma:defn}, respectively, for four
different values of the bosonic bath exponent $s$. Parentheses surround the
estimated nonsystematic error in the last digit.}
\begin{ruledtabular}
\begin{tabular}{lllll}
\multicolumn{1}{c}{$s$} & \multicolumn{1}{c}{$\beta$} &
  \multicolumn{1}{c}{$1/\delta$} & \multicolumn{1}{c}{$x$} &
  \multicolumn{1}{c}{$\gamma$} \\ \hline \\[-2ex]
0.2 & 2.0005(3) & 0.6673(1) & 0.1997(2) & 0.997(4) \\
0.4 & 0.7568(2) & 0.4283(2) & 0.4002(4) & 1.0117(6) \\
0.6 & 0.3923(1) & 0.2501(7) & 0.600(2)  & 1.1805(5) \\
0.8 & 0.2130(1) & 0.1111(1) & 0.800(2)  & 1.703(3)
\end{tabular}
\end{ruledtabular}
\end{table}

\subsubsection{Dynamical local charge susceptibility}
\label{subsec:dynamical-response}

The dynamical local charge susceptibility is
\begin{equation}
\chi_{c,\loc}(\omega,T)=i \! \int^{\infty}_{0} \!\!\! dt\; e^{-i\omega t} \:
  \bigl\langle\!\bigl\langle [ \n_d(t)-1, \,
  \n_d(0)-1 ] \bigr\rangle\!\bigr\rangle .
\end{equation}
Its imaginary part $\chi_{c,\loc}^{\prime\prime}$ can be calculated within
the NRG as
\begin{multline}
\label{dynamical-chi}
\chi_{c,\loc}^{\prime\prime}(\omega,T) = \frac{\pi}{Z(T)} \sum_{m,m'}
  \bigl| \bigl\langle m'|\n_d-1|m \bigr\rangle\bigr|^2 \\
\times \bigl( e^{-E_{m'}/T} - e^{-E_m/T} \bigl) \delta(\omega-E_{m'}+E_m) .
\end{multline}
Here, $|m\rangle$ is a many-body eigenstate with energy $E_m$,
and $Z(T)=\sum_m e^{-E_m/T}$ is the partition function.
Equation \eqref{dynamical-chi} produces a discrete set of delta-function peaks
that must be broadened to recover a continuous spectrum. Following standard
procedure,\cite{Sakai:89+Bulla:01} we employ Gaussian broadening of delta
functions on a logarithmic scale:
\begin{equation}
\label{broadening}
\delta(|\omega|\!-\!|\Delta E|) \to
  \frac{e^{-b^2/4}}{\sqrt{\pi}\,b\,|\Delta E|} \:
  \exp\left[-\frac{(\ln|\omega|-\ln|\Delta E|)^2}{b^2}\right],
\end{equation}
with the choice of the broadening width $b=0.5\ln\Lambda$.

\begin{figure}
\centering
\includegraphics[angle=270,width=0.9\columnwidth]{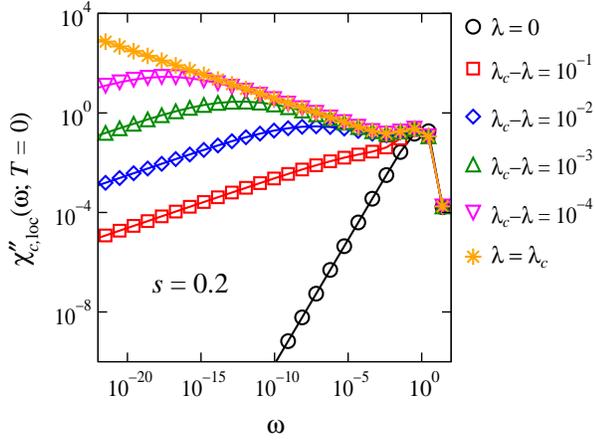}
\caption{\label{fig:dynamical-chi}
(Color online)
Imaginary part of the dynamical local charge susceptibility
$\chi_{c,\loc}^{\prime\prime}(\omega;T=0)$ vs frequency $\omega$ for
$s=0.2$, $U=-2\epsilon_d=0.1$, $\Gamma=0.5$ (see Footnote
\protect\onlinecite{large-Gammas}), $\Lambda=9$, $N_s=500$, $N_b=8$,
and different \eb\ couplings $\lambda<\lambda_c$ on the Kondo side of the
critical point, which is located at $\lambda_c\simeq0.53008$. As
$\lambda\rightarrow\lambda_c^-$, $\chi_{c,\loc}^{\prime\prime}(\omega;T=0)$
follows the quantum critical form [Eq.\ \protect\eqref{dynamical-chi-critical}]
for $T_*\ll\omega\ll T_{\K}$, where $T_{\K}$ is the Kondo scale of the
pure-fermionic ($\lambda=0$) problem.}
\end{figure}

\begin{figure}
\centering
\includegraphics[angle=270,width=0.85\columnwidth]{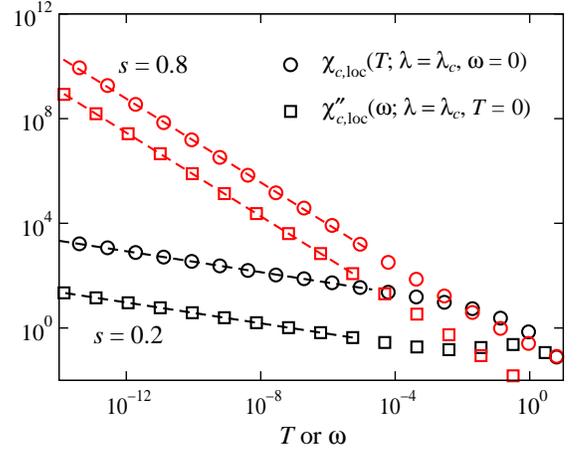}
\caption{\label{fig:x=y}
(Color online) Critical static and dynamical response:
$\chi_{c,\loc}(T;\lambda=\lambda_c,\omega=0)$ vs $T$ (circles) and
$\chi_{c,\loc}^{\prime\prime}(\omega;\lambda=\lambda_c,T=0)$ vs $\omega$
(squares) for two representative bath exponents $s=0.2$ and $s=0.8$. All other
parameters are as in Fig.\ \protect\ref{fig:W_d,L}. The equality of the slopes
of the static and dynamical charge susceptibilities for a given bath exponent
$s$ indicates that the corresponding critical exponents satisfy $x=y$.}
\end{figure}

(a) \textit{Zero temperature.}
Figure~\ref{fig:dynamical-chi} plots $\chi_{c,\loc}^{\prime\prime}(\omega;T=0)$
vs $\omega$ for bath exponent $s=0.2$ and a series of \eb\ couplings
$\lambda<\lambda_c$.
Whereas $\chi_{c,\loc}^{\prime\prime}(\omega;\lambda=0,T=0) \propto \omega$ for
$|\omega|\ll T_{\K}$ (the usual Kondo result), we find that
$\chi_{c,\loc}^{\prime\prime}(\omega;0<\lambda<\lambda_c,T=0) \propto
|\omega|^s\sgn(\omega)$ as $\omega\rightarrow 0$, corresponding to a
long-time relaxation behavior $\chi_{c,\loc}(t)\propto t^{-(1+s)}$.
Precisely at the critical \eb\ coupling,
\begin{equation}
\label{dynamical-chi-critical}
\chi_{c,\loc}^{\prime\prime}(\omega;\lambda=\lambda_c,T=0)\propto
|\omega|^{-y}\sgn(\omega) \quad \text{for } \omega\ll T_{\K} .
\end{equation}
Figure \ref{fig:x=y} shows
$\chi_{c,\loc}^{\prime\prime}(\omega;\lambda=\lambda_c,T=0)$ vs $\omega$
and $\chi_{c,\loc}(T;\lambda=\lambda_c,\omega=0)$ vs $T$ for
representative bosonic bath exponents $s=0.2$ and $s=0.8$.
These and all other data that we have obtained are consistent with the
relation
\begin{equation}
\label{x=y}
x=y=s \qquad \text{for } 0<s<1.
\end{equation}

For small deviations from the critical coupling,
$\chi_{c,\loc}^{\prime\prime}(\omega;T=0)$ exhibits the
critical behavior of Eq.\ \eqref{dynamical-chi-critical} over the range
$T_*\ll |\omega| \ll T_{\K}$, where $T_*$ is identical (up to a constant
multiplicative factor) to the crossover scale defined in
Sec.\ \ref{subsec:T*} that vanishes at the quantum critical point according
to Eq.\ \eqref{nu:defn}.

\begin{figure}
\centering
\includegraphics[angle=270,width=0.85\columnwidth]{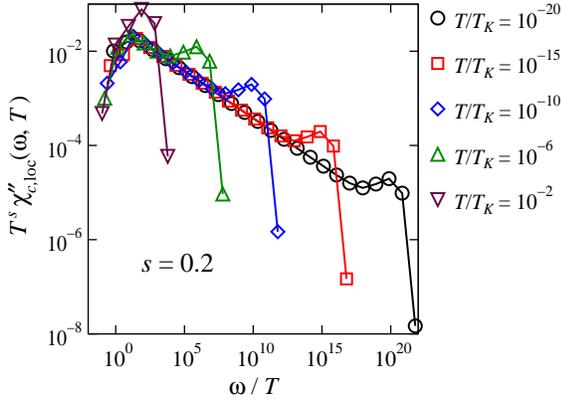}
\caption{\label{fig:wTscaling}
(Color online)
Scaling with $\omega/T$ of the imaginary part of the dynamical local charge
susceptibility $\chi_{c,\loc}^{\prime\prime}(\omega,T)$ at the critical \eb\
coupling $\lambda_c\simeq0.53008$ for $s=0.2$, $U=-2\epsilon_d=0.1$,
$\Gamma=0.5$ (see footnote \protect\onlinecite{large-Gammas}), $\Lambda=9$,
$N_s=500$, $N_b=8$, and different temperatures $T\ll T_{K}=0.425$.}
\end{figure}

(b) \textit{Finite temperatures.}
Equation \eqref{x=y} is consistent with the presence of $\omega/T$ scaling in
the dynamical local charge susceptibility at the quantum critical point, viz
\begin{equation}
\chi_{c,\loc}^{\prime\prime}(\omega,T;\lambda=\lambda_c)=T^{-s}\Psi_s(\omega/T).
\end{equation}
Figure~\ref{fig:wTscaling} shows the collapse of data for
$\chi_{c,\loc}^{\prime\prime}(\omega,T;\lambda=\lambda_c)$ onto a single
function of $\omega/T$ within the critical regime. The Kondo temperature
$T_{\K}$ of the Anderson model obtained by setting $\lambda=0$ serves as a
nonuniversal high-frequency cutoff on the critical behavior; the curves have
a common form for $\omega/T\ll T_{\K}/T$. It should be noted that the NRG
method is unreliable\cite{Costi:94,Ingersent:02} for $|\omega|\lesssim T$,
preventing demonstration of complete $\omega/T$ scaling.

Both the hyperscaling of the static critical exponents and what seems
to be $\omega/T$ scaling of the dynamical susceptibility are consistent
with the QPT between the Kondo and localized phases taking place at
an interacting critical point below its upper critical dimension.

\subsection{Impurity spectral function}
\label{subsec:specfun}

We now turn to discussion of the impurity spectral function
$A_{\sigma}(\omega,T)=-\pi^{-1} \text{Im}\,G_{d\sigma}(\omega, T)$, where
the retarded impurity Green's function is
\begin{equation}
G_{d\sigma}(\omega,T)=-i \! \int_0^{\infty} \!\!\! dt \, e^{i\omega t} \:
 \bigl\langle\!\bigl\langle \bigl[ d_{\sigma}^{\pdag}(t), \,
 d_{\sigma}^{\dag}(0) \bigr]_{+} \bigr\rangle\!\bigr\rangle \, .
\end{equation}
The spectral function can be calculated within the NRG using the
formulation
\begin{multline}
\label{specfun}
A_{\sigma}(\omega, T) = \frac{1}{Z(T)} \sum_{m,m'}
  \bigl|\bigl\langle m'|d_{\sigma}^{\dag}|m \bigr\rangle \bigr|^2 \\
  \bigl( e^{-E_{m'}/T} + e^{-E_m/T} \bigr)
\times \delta(\omega-E_{m'}+E_m),
\end{multline}
where the notation is the same as in Eq.\ \eqref{dynamical-chi}.
To recover a continuous spectrum, we have again applied Eq.\
\eqref{broadening} to the delta-function output of Eq.\ \eqref{specfun},
choosing the broadening factor $b=0.55\ln\Lambda$ that best satisfies the
Fermi-liquid result $A_{\sigma}(\omega=0, T=0)=1/\pi\Gamma$ for
the Anderson model. In order to achieve satisfactory results,
we find it necessary to work with a smaller discretization parameter
($\Lambda=3$ instead of the value $\Lambda=9$ employed for all the
quantities reported above) and to retain more states ($N_s = 1200$ rather
than the 500 that typically suffices). Since the spectral functions shown
below are all spin-independent, we henceforth drop the index $\sigma$ on
$A_{\sigma}$. For the particle-hole-symmetric model considered in this
section, the spectral function is symmetric about $\omega=0$.

\begin{figure}
\centering
\includegraphics[angle=270,width=0.85\columnwidth]{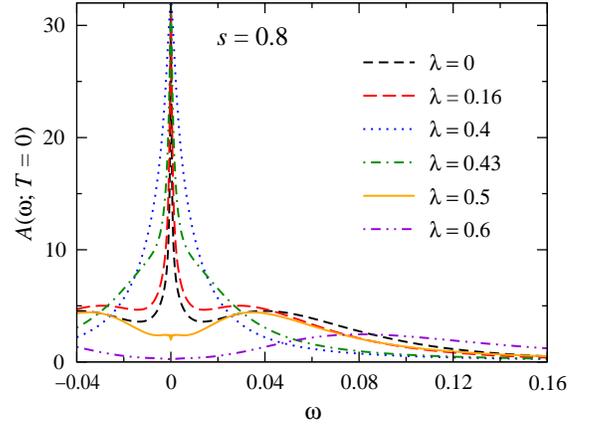}
\caption{\label{fig:specfun}
(Color online)
Impurity spectral function $A(\omega;T=0)$ vs frequency $\omega$ for
$s=0.8$, $U=-2\epsilon_d=0.1$, $\Gamma=0.01$, $\Lambda=3$, $N_s=1200$,
$N_b=8$, and different values of the \eb\ coupling $\lambda$.
For these parameters, $U_{\eff}$ defined in
Eq.\ \protect\eqref{Ueff0:NRG} changes sign at $\lambda_{c0}\simeq 0.369$
and the critical coupling is $\lambda_c\simeq 0.474$.}
\end{figure}

Figure \ref{fig:specfun} plots $A_{\sigma}(\omega;T=0)$ vs $\omega$
for $s=0.8$ and a series of $\lambda$ values. For $\lambda=0$, we recover
the spectral function of the Anderson model, featuring a
narrow Kondo resonance centered at zero frequency and broad Hubbard
satellite bands centered around $\omega=\pm\frac{1}{2} U$.
Increasing the \eb\ coupling from zero has two initial effects---a
displacement of the Hubbard bands to smaller frequencies, and a
broadening of the low-energy Kondo resonance---that can both be attributed
to the boson-induced renormalization of the Coulomb interaction described in
Eq.\ \eqref{Ueff}.

\begin{figure}
\centering
\includegraphics[angle=270,width=0.85\columnwidth]{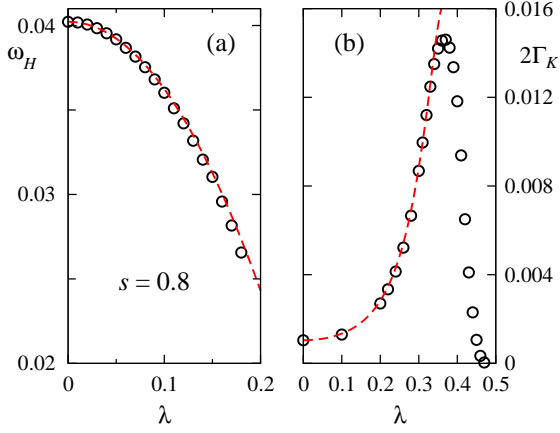}
\caption{\label{fig:omega_H+Gamma_K}
(Color online)
Variation with \eb\ coupling $\lambda$ of two characteristic energy
scales extracted from the zero-temperature impurity spectral function.
All parameters except $\lambda$ are the same as in
Fig.\ \protect\ref{fig:specfun}.
(a) Location $\omega_H$ of the upper Hubbard peak. The dashed line shows
$\omega_H=0.4 U -\lambda^2/(\pi s)$.
(b) Kondo resonance width (full width at half height) $2\Gamma_{\K}$.
The dashed line, representing the prediction of
Eq.\ \protect\eqref{Gamma_K-vs-Ueff} with $C_{\K} = 0.82$ and with
$\tilde{U}^{\NRG}$ in Eq.\ \protect\eqref{Ueff:NRG} evaluated at
$E=U/2=|\epsilon_d|$, fits the data over almost the entire range
$0\le\lambda<\lambda_{c0}\simeq 0.369$.}
\end{figure}

We expect the Hubbard peak locations to obey
$\omega_H\simeq\pm\frac{1}{2}U_{\eff}$ for
$0 \le \lambda \ll \lambda_{c0}$. However, the peak locations plotted in
Fig.\ \ref{fig:omega_H+Gamma_K}(a) are better fitted by
$|\omega_H|= 0.4 U - \lambda^2/(\pi s)$, which (given the
discretization and truncation effects discussed in Sec.\ \ref{subsec:Gamma=0})
appears to represent a stronger bosonic renormalization than that predicted by
$|\omega_H|=\frac{1}{2}U_{\eff}$. We believe that this discrepancy arises
primarily from the rapid broadening of the Kondo resonance with increasing
$\lambda$, which shifts the local maximum of the combined spectral function
(the sum of the Kondo resonance plus Hubbard satellite bands) to a
frequency smaller in magnitude than the central frequency of the Hubbard peak
by itself.

The width $2\Gamma_{\K}$ of the Kondo resonance, plotted in
Fig.\ \ref{fig:omega_H+Gamma_K}(b), proves to be equal (up to a multiplicative
constant) to the crossover scale $T_*$ defined in Sec.\ \ref{subsec:T*}.
For $\lambda\lesssim\lambda_{c0}$, the variation in both scales is well
described by the replacement of $U$ in the expression\cite{Krishna-murthy:80}
for the Kondo temperature of the symmetric Anderson model by
$\tilde{U}^{\NRG}(U/2)$ [given by Eq.\ \eqref{Ueff:NRG}], the effective
Coulomb interaction on entry to the local-moment regime. The dashed line
in Fig.\ \ref{fig:omega_H+Gamma_K}(b) shows that the resulting formula,
\begin{equation}
\label{Gamma_K-vs-Ueff}
\Gamma_{\K} = C_{\K} \sqrt{\frac{8\tilde{U}^{\NRG}\Gamma}{\pi A_{\Lambda}}}
   \exp\left(-\frac{\pi A_{\Lambda}\tilde{U}^{\NRG}}{8\Gamma}\right) ,
\end{equation}
where $A_{\Lambda}$ is defined in Eq.\ \eqref{A_Lambda}, provides an
excellent description of $\Gamma_{\K}$ over almost the entire range
$0\le\lambda<\lambda_{c0}\simeq 0.369$. This echoes the finding in the
Anderson-Holstein model that a weak \eb\ coupling serves primarily to
reduce the impurity on-site repulsion, leading to an increase in the
Kondo scale.\cite{Cornaglia:04+05}

\begin{figure}
\centering
\includegraphics[angle=270,width=0.85\columnwidth]{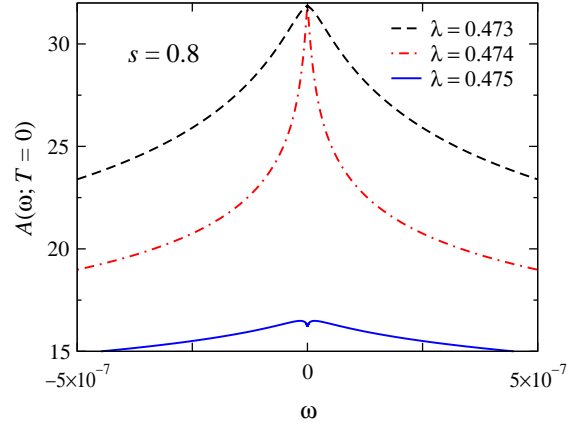}
\caption{\label{fig:specfun:pinning}
(Color online)
Detail of the impurity spectral function $A(\omega;T=0)$ around frequency
$\omega=0$ for $s=0.8$, $U=-2\epsilon_d=0.1$, $\Gamma=0.01$, $\Lambda=3$,
$N_s=1600$, $N_b=8$, and different \eb\ couplings $\lambda$ straddling the
critical value $\lambda_c\simeq 0.47458$. For $\lambda\le\lambda_c$,
$A(\omega;T=0)$ is pinned to the value predicted by Fermi-liquid theory. For
$\lambda>\lambda_c$, the Kondo resonance disappears, leaving a pair of
low-energy peaks centered at $|\omega|$ of order the crossover temperature
$T_*$ ($\simeq \! 1.4\times 10^{-8}$ for $\lambda\!=\!0.475$).}
\end{figure}

Once the \eb\ coupling exceeds $\lambda_{c0}$, further increase in
$\lambda$ leads to suppression of the Hubbard peaks (e.g., see the
curves for $\lambda=0.4$ and $\lambda=0.43$ in Fig.\ \ref{fig:specfun}) and to a rapid
narrowing of the Kondo resonance [see Fig.\ \ref{fig:omega_H+Gamma_K}(b)].
In the Anderson-Holstein model, the Kondo scale remains nonzero---although
exponentially reduced---for arbitrarily large \eb\
couplings.\cite{Cornaglia:04+05} In the charge-coupled BFA model, by contrast,
the Kondo peak collapses and $\Gamma_{\K}$ extrapolates to zero as $\lambda$
approaches its critical value $\lambda_c$. As shown in
Fig.\ \ref{fig:specfun:pinning}, the central peak remains pinned to
the Fermi-liquid result $A(\omega=0,T=0)=1/\pi \Gamma$ even as the
peak width vanishes for $\lambda\to\lambda_c^-$.

In the localized phase ($\lambda > \lambda_c$), there is no vestige of the
Kondo resonance, but high-energy Hubbard-like peaks reappear; see the curves
for $\lambda = 0.5$ and 0.6 in Fig.\ \ref{fig:specfun}.
In addition, there is a pair of low-energy peaks centered at
$\omega\simeq \pm T_*$, as shown in Fig.\ \ref{fig:specfun:pinning}.

\subsection{Spin-Kondo to charge-Kondo crossover}
\label{subsec:crossover}

\begin{figure}
\centering
\includegraphics[angle=270,width=0.75\columnwidth]{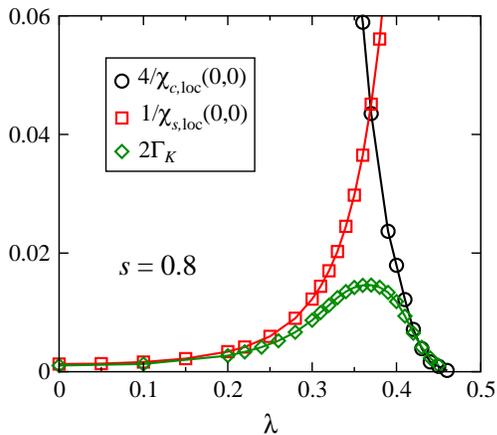}
\caption{\label{fig:Tscale}
(Color online)
Variation with \eb\ coupling $\lambda<\lambda_c$ of the Kondo resonance
width $2\Gamma_{\K}$, the inverse static local spin susceptibility
$1/\chi_{s,\loc}(\omega=0,T=0)$, and the inverse static local charge
susceptibility $4/\chi_{c,\loc}(\omega=0,T=0)$. The results shown are for
$s=0.8$, $U=-2\epsilon_d=0.1$, $\Gamma=0.01$, $\Lambda=3$, $N_s=1200$, and
$N_b=8$. For the calculation of the static local spin susceptibility via
Eq.\ \protect\eqref{static-chi_s}, the total spin $S$ is not a good quantum
number, so $N_s$ specifies the number of $(S_z,Q)$ states retained after each
iteration.}
\end{figure}

Based on the analysis of the zero-hybridization limit presented in
Sec. \ref{subsec:Gamma=0}, one expects spin fluctuations to dominate
the impurity behavior in the region $\lambda \ll \lambda_{c0}$, but charge
fluctuations to be dominant for $\lambda_{c0} \ll \lambda < \lambda_c$.
This picture is supported by the behaviors of the thermodynamic
susceptibilities discussed in Sec.\ \ref{subsec:thermodynamics}.
The evolution from a spin-Kondo effect to a charge-Kondo effect can also
be probed by comparing the static local charge susceptibility
[Eq.\ \eqref{static-chi}] with its spin counterpart
\begin{equation}
\label{static-chi_s}
\chi_{s,\loc}(T; \omega=0) = -\lim_{h\to 0}
  \frac{\langle\!\langle \n_{d\uparrow} -
     \n_{d\downarrow}\rangle\!\rangle}{2h},
\end{equation}
where $h$ is a local magnetic field that enters an additional Hamiltonian
term
\begin{equation}
\label{H_s,loc}
\H_{s,\loc} = \frac{h}{2} \, (\n_{d\uparrow}-\n_{d\downarrow}) .
\end{equation}
In particular, characteristic energy scales for the spin and charge
Kondo effects are expected to be $1/\chi_{s,\loc}(\omega=0,T=0)$ and
$4/\chi_{c,\loc}(\omega=0,T=0)$, respectively [where the factor of $4$
accounts for the difference in conventions that $\phi$ couples to $\n_d-1$,
whereas $h$ couples to $(\n_{d\uparrow}-\n_{d\downarrow})/2$].
Figure \ref{fig:Tscale} plots the $\lambda$ dependence of these quantities
for the parameter set illustrated in Figs.\ \ref{fig:specfun} and
\ref{fig:omega_H+Gamma_K}. The Kondo resonance width $2\Gamma_{\K}$ crosses over
from paralleling $1/\chi_{s,\loc}(0,0)$ for small $\lambda$ to loosely
tracking\cite{tracking} $4/\chi_{c,\loc}(0,0)$ as $\lambda$ approaches
$\lambda_c$. In the intermediate region near $\lambda=\lambda_{c0}$,
$2\Gamma_{\K}$ is much smaller than either inverse static susceptibility,
indicating that the Kondo effect has mixed spin and charge character.

\begin{figure}
\centering
\includegraphics[angle=270,width=0.8\columnwidth]{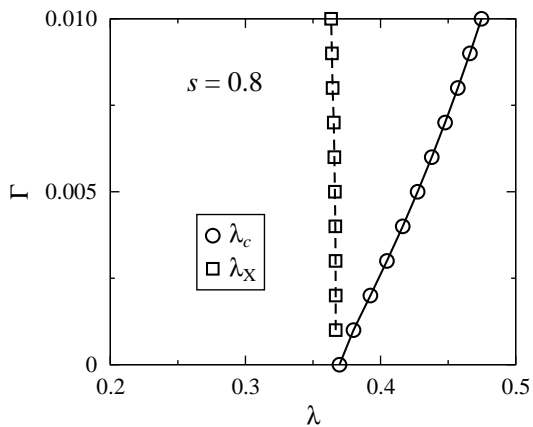}
\caption{\label{fig:critandcross}
Phase boundary $\lambda_c(\Gamma)$ and crossover boundary
$\lambda_{\text{X}}(\Gamma)$ (defined in the text) for $s=0.8$,
$U=-2\epsilon_d=0.1$, $\Lambda=3$, $N_s=1200$, and $N_b=8$.
The data are consistent with the schematic phase diagram
shown in Fig.\ \ref{fig:phase-diagram}.}
\end{figure}

Figure \ref{fig:critandcross} presents a $\lambda$-$\Gamma$ phase diagram for
$s=0.8$ and fixed $U=-2\epsilon_d$, showing data points along the phase
boundary $\lambda=\lambda_c(\Gamma)$ and along the crossover boundary
$\lambda=\lambda_{\text{X}}(\Gamma)$, defined as the \eb\ coupling at which
the Kondo resonance width $2\Gamma_{\K}$ is maximal for the given $\Gamma$.
The fact that the latter line rises almost vertically from $\lambda=\lambda_{c0}$
at $\Gamma=0$ provides further confirmation of the picture of a crossover from
a spin-Kondo effect to a charge-Kondo effect resulting from the change in the
sign of $U_{\eff}$, and establishes the validity of the schematic phase diagram
(Fig.\ \ref{fig:phase-diagram}) presented in the introduction to this section.

\section{Results: Symmetric Model with Ohmic dissipation}
\label{sec:ohmic}

This section presents results for Hamiltonian \eqref{H_CCBFA} with
$U=-2\epsilon_d > 0$ and an Ohmic bath (i.e., $s=1$). We first discuss the
behavior of the static local charge susceptibility. We show that, in
contrast with the sub-Ohmic case $0<s<1$, the crossover scale vanishes in
exponential (rather than power-law) fashion as the \eb\ coupling
approaches its critical value from below, and there is no small energy
scale observed on the localized side of the transition. Therefore, the
QPT for the Ohmic case is
of Kosterlitz-Thouless type. At the end of the section, we study the
effects of the \eb\ coupling on the impurity spectral function.

\subsection{Fixed points and thermodynamic susceptibilities}
\label{subsec:thermodynamics:s=1}

\begin{figure}
\centering
\includegraphics[width=0.75\columnwidth]{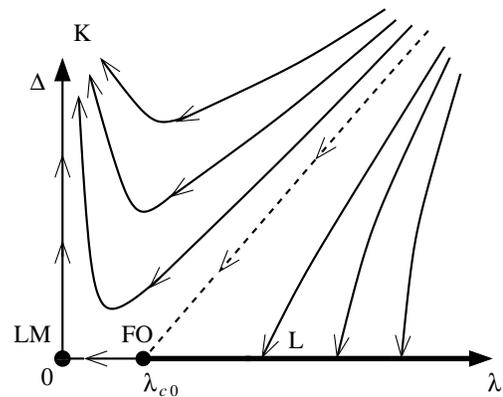}
\caption{\label{fig:RGflows:s=1}
Schematic renormalization-group flows on the $\lambda$-$\Delta$ plane for the
symmetric model with bath exponent $s=1$. Trajectories represent the flow of
the couplings $\lambda$ entering Eq.\ \protect\eqref{H_impbath:NRG} and
$\Delta$ defined in Eq.\ \protect\eqref{Delta:defn} under decrease in the
high-energy cutoffs on the conduction band and the bosonic bath. A separatrix
(dashed line) forms the boundary between the basins of attraction of the Kondo
fixed point (K) and a line of localized fixed points (L). Flow along the
separatrix is toward the free-orbital fixed point (FO) located at
$\lambda=\lambda_{c0}$. For $\Delta=0$ only, there is flow away from FO
toward the local-moment fixed point (LM) at $\lambda=0$.}
\end{figure}

Figure \ref{fig:RGflows:s=1} plots the schematic renormalization-group flows
for a symmetric impurity coupled to an Ohmic bath. The flows within the
Kondo basin of attraction are qualitatively very similar to those for
the sub-Ohmic case depicted in Fig.\ \ref{fig:RGflows}. In the localized
regime, however, the \eb\ coupling flows not to $\lambda = \infty$, but
rather to a finite limiting value that varies continuously with the bare
values of $\lambda$ and $\Gamma$. What is shown as a line of fixed points
in Fig.\ \ref{fig:RGflows:s=1} is really a plane of fixed points described by
$\H_{\LC}$ [Eq.\ \eqref{H_LC}] with effective couplings $\lambda>\lambda_{c0}$,
$W_p=0$, and $0\le W_d <\infty$. Another important departure from the sub-Ohmic
case is that for $s=1$ there is no longer a distinct critical point reached by
flow along the separatrix from the free-orbital fixed point; rather these two
fixed points merge as $s\to 1^-$, leaving a critical endpoint at
$\lambda=\lambda_{c0}$, $\Delta=0$. Strictly, this is a line of critical
endpoints described by $\H_{\LC}$ [Eq.\ \eqref{H_LC}] with effective couplings
$\lambda=\lambda_{c0}$, $W_p=0$, and $0\le W_d <\infty$. For a fixed bare value
of $\Gamma$, the endpoint value of $W_d$ is just the limit of the localized
fixed-point value of $W_d$ as the bare coupling $\lambda$ approaches the phase
boundary $\lambda_c(\Gamma)$.

The behaviors of the static impurity spin and charge susceptibilities are
qualitatively very similar to those for a sub-Ohmic bath, as discussed in
Sec.\ \ref{subsec:thermodynamics}. The only significant difference is that
for $s=1$, $\lim_{T\to 0}T\chi_{c,\imp}(T)$ undergoes a discontinuous jump
from its value of 0 for $\lambda\le\lambda_c$ to a nonzero value for
$\lambda=\lambda_c^+$. This jump can be understood through
Eqs.\ \eqref{delta_0-vs-W_d} and \eqref{Tchi_c-vs-delta_0} as a consequence of
the fact that $W_d$ does not diverge on approach to the critical
coupling.

\subsection{Static local charge susceptibility and crossover scale}
\label{subsec:static-response:s=1}

\begin{figure}
\centering
\includegraphics[angle=270,width=0.9\columnwidth]{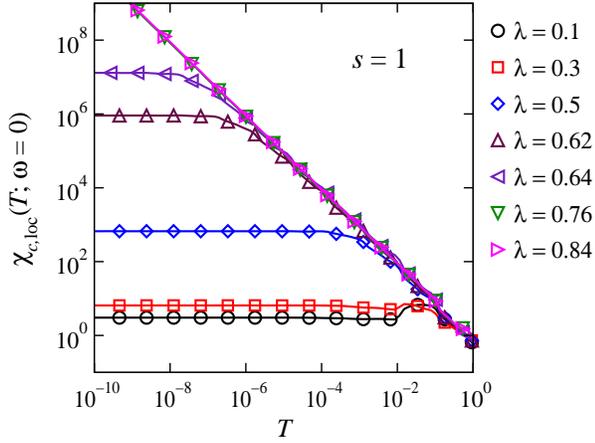}
\caption{\label{fig:static-chi:s=1}
(Color online) Static local charge susceptibility
$\chi_{c,\loc}(T;\omega=0)$ vs temperature $T$ for $s=1$,
$U=-2\epsilon_d=0.1$, $\Gamma=0.01$, $\Lambda=9$, $N_s=800$, $N_b=12$,
and different \eb\ couplings $\lambda$. On the Kondo side of the
QPT ($\lambda<\lambda_c\simeq 0.726$), there is a
clear crossover from quantum-critical to screened behavior around the
renormalized Kondo temperature $T_* = 4/\chi_{c,\loc}(\omega=T=0)$.
No such crossover is evident on the localized side ($\lambda>\lambda_c$).}
\end{figure}

Figure~\ref{fig:static-chi:s=1} is a logarithmic plot of the static local
charge susceptibility $\chi_{c,\loc}(T;\omega=0)$ vs temperature $T$ for
different \eb\ couplings $\lambda$. On the Kondo side of the phase boundary,
$\chi_{c,\loc}(T;\omega=0)$ is proportional to $1/T$ at high temperatures,
but levels off for $T\lesssim T_*$. We find it convenient to \textit{define}
\begin{equation}
\label{T*:Kondo:s=1}
T_* = 4/\chi_{c,\loc}(\omega=T=0) \quad \text{for } \lambda\to\lambda_c^-,
\end{equation}
thereby removing the ambiguity in the definition of the crossover iteration
$N_*$ (see Sec.\ \ref{subsec:T*}) on the Kondo side of the $s=1$ quantum
phase transition.

\begin{figure}
\centering
\includegraphics[angle=270,width=0.8\columnwidth]{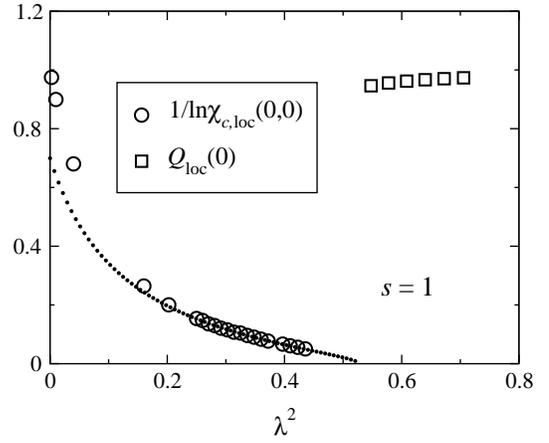}
\caption{\label{fig:T*+Qloc:s=1}
Variation with \eb\ coupling $\lambda$ of the local charge
susceptibility $\chi_{c,\loc}(\omega=T=0)$ in the Kondo phase
$\lambda<\lambda_c\simeq 0.726$ and of the order parameter
$\lim_{\phi\to 0^-} Q_{\loc}(\phi;T=0)$ in the localized phase
$\lambda>\lambda_c$, for $s=1$, $U=-2\epsilon_d=0.1$, $\Gamma=0.01$,
$\Lambda=9$, $N_s=800$, and $N_b=12$. The dotted line shows a fit of
the susceptibility data using Eqs.\ \protect\eqref{T*:Kondo:s=1} and
\protect\eqref{T*:s=1}.}
\end{figure}

For $\lambda\rightarrow\lambda^{-}_c$,
the crossover scale vanishes according to (see Fig.\ \ref{fig:T*+Qloc:s=1})
\begin{equation}
\label{T*:s=1}
T_* \propto \exp\left[ -\frac{C_*}{\sqrt{1-(\lambda/\lambda_c)^2}} \right] .
\end{equation}

In the localized phase, $\chi_{c,\loc}(T;\omega=0)$ satisfies
Eq.\ \eqref{Curie-law} over the entire temperature range $T\ll U$.
Since the critical and localized fixed points share the same temperature
variation, no crossover scale can be identified on the localized side of the
phase boundary. Moreover, the order parameter
$\lim_{\phi\to 0} Q_{\loc}(\phi;T=0)$ does not vanish continuously as
$\lambda\rightarrow \lambda_c^+$, but rather undergoes a discontinuous jump
at the transition, as shown in Fig.\ \ref{fig:T*+Qloc:s=1}. The magnitude
of this jump is nonuniversal, being related via Eq.\ \eqref{Q_loc-vs-W_d}
to the value of $W_d$ at the critical endpoint.

The properties described above are analogous to those of the Kondo model
[Eq.\ \eqref{H_K}] at the transition between the Kondo-screened phase
(reached for $J_{\perp} \ne 0$ and $J_z > -|J_{\perp}|$) and the
local-moment phase (reached for $J_z\le -|J_{\perp}|$).
Such behaviors are characteristic of a Kosterlitz-Thouless type of QPT.

\subsection{Impurity spectral function}
\label{subsec:specfun:s=1}

\begin{figure}
\centering
\includegraphics[angle=270,width=0.8\columnwidth]{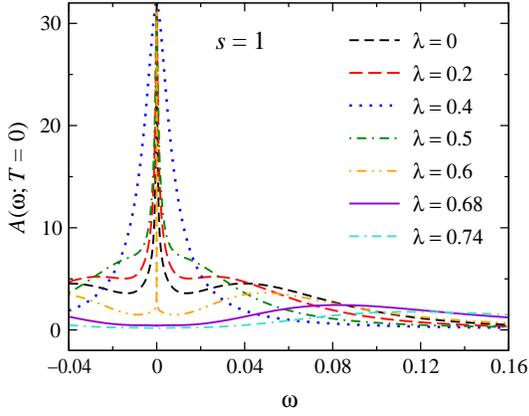}
\caption{
\label{fig:specfun:s=1}
(Color online)
Impurity spectral function $A(\omega;T=0)$ vs
$\omega$ for $s=1$, $U=-2\epsilon_d=0.1$, $\Gamma=0.01$, $\Lambda=3$,
$N_s=1200$, $N_b=12$, and different values of the \eb\ coupling $\lambda$.
For these parameters, $U_{\eff}$ [Eq.\ \protect\eqref{Ueff0:NRG}] changes
sign at $\lambda_{c0}\simeq 0.413$ and the critical coupling is
$\lambda_c\simeq 0.669$.}
\end{figure}

\begin{figure}
\centering
\includegraphics[angle=270,width=0.85\columnwidth]{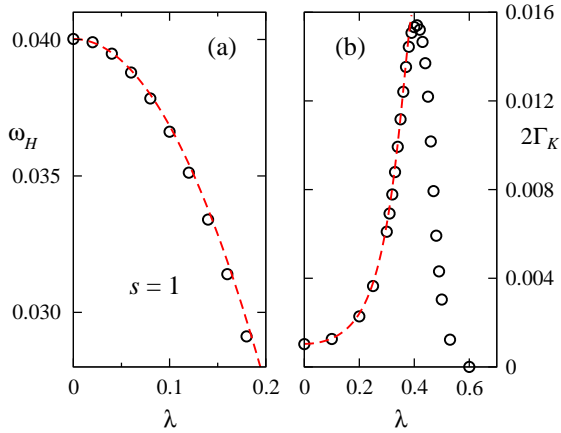}
\caption{\label{fig:omega_H+Gamma_K:s=1}
(Color online)
Variation with \eb\ coupling $\lambda$ of two characteristic energy
scales extracted from the zero-temperature impurity spectral function.
All parameters except $\lambda$ are the same as in
Fig.\ \protect\ref{fig:specfun:s=1}.
(a) Location $\omega_H$ of the upper Hubbard peak. The dashed line shows
$\omega_H(\lambda)=0.4 U -\lambda^2/\pi$.
(b) Kondo resonance width (full width at half height) $2\Gamma_{\K}$.
The dashed line, representing the prediction of
Eq.\ \protect\eqref{Gamma_K-vs-Ueff} with $C_{\K} = 0.82$ and with
$\tilde{U}^{\NRG}$ in Eq.\ \protect\eqref{Ueff:NRG} evaluated at
$E=U/2=|\epsilon_d|$, fits the data over almost the entire range
$0\le\lambda<\lambda_{c0}\simeq 0.413$.}
\end{figure}

Figure \ref{fig:specfun:s=1} shows the impurity spectral function
$A(\omega;T=0)$ for an Ohmic bath. The behavior in the Kondo phase is similar
to that in the sub-Ohmic case discussed in Sec.\ \ref{subsec:specfun}: As the
\eb\ coupling $\lambda$ increases from zero, the Hubbard satellite bands
are initially displaced to smaller frequencies according to $\omega_H\simeq
\pm\frac{1}{2}U_{\eff}$ [Fig.\ \ref{fig:omega_H+Gamma_K:s=1}(a)], while the
width $2\Gamma_{\K}$ of the Kondo resonance [Fig.\
\ref{fig:omega_H+Gamma_K:s=1}(b)] first rises before falling sharply on
approach to $\lambda=\lambda_c$. Just as for $0<s<1$, the variation in
$\Gamma_{\K}$ for $\lambda\lesssim\lambda_{c0}$ is well described
by Eq.\ \eqref{Gamma_K-vs-Ueff} with $\tilde{U}^{\NRG}$ [Eq.\ \eqref{Ueff:NRG}]
evaluated at $E=U/2$. Throughout the Kondo phase, $A(\omega=T=0)$ remains pinned
at its Fermi-liquid value $1/\pi\Gamma$.

\begin{figure}
\centering
\includegraphics[angle=270,width=0.8\columnwidth]{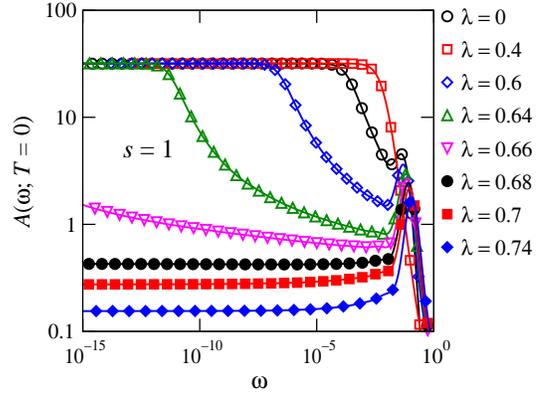}
\caption{\label{fig:logspecfun:s=1}
(Color online)
Impurity spectral function $A(\omega;T=0)$ vs frequency $\omega$ on a
logarithmic scale for $s=1$, $U=-2\epsilon_d=0.1$,
$\Gamma=0.01$, $\Lambda=3$, $N_s=1200$, $N_b=12$, and different \eb\
couplings $\lambda$. For $\lambda<\lambda_c\simeq 0.669$, the behavior is
similar to that found for $0<s<1$. However, for $\lambda\ge\lambda_c$,
the spectral function is essentially featureless below the energy scale
$\frac{1}{2} |U_{\eff}|$ of the Hubbard peaks.}
\end{figure}

For $\lambda\ge \lambda_c$, however, the behavior of the spectral function
is quite different for $s=1$ than for $0<s<1$. In the sub-Ohmic case,
the Kondo-phase pinning extends to the quantum critical point, i.e.,
$\pi\Gamma A(\omega=T=0,\lambda=\lambda_c)=1$, while in the localized
phase peaks appear at $\omega\simeq \pm T_*$.
Figure \ref{fig:logspecfun:s=1} shows that the Ohmic spectral function instead
satisfies $\pi\Gamma A(\omega=T=0,\lambda=\lambda_c)<1$, and exhibits no
feature in the localized phase at energy scales much smaller than
$\frac{1}{2} |U_{\eff}|$.

\section{Results: Asymmetric Model}
\label{sec:asymm}

Sections \ref{sec:sub-ohmic} and \ref{sec:ohmic} focused exclusively on
results for a symmetric impurity satisfying $\epsilon_d = -U/2$ in
Eq. \eqref{H_imp} or, equivalently, $\delta_d = 0$ in
Eq.\ \eqref{H_imp:delta}. We now turn to the general situation of an
asymmetric impurity, starting with the sub-Ohmic case $0<s<1$.

For $\delta_d\ne 0$ and small, nonzero values of $\lambda$, one expects
the fermionic sector of the charge-coupled BFA model to behave in
essentially the same manner as in the asymmetric Anderson model (reviewed
in Sec.\ \ref{subsec:lambda=0}), with the exception that the effective
value of the Coulomb interaction $U$ will be reduced by the coupling to
the bosonic bath. At temperatures well below $T_{\K}$, there will be no
further renormalization of the electronic degrees of freedom, the system will
exhibit quasiparticle excitations described by $\H_{\SC}^{\NRG}$ in
Eq.\ \eqref{H_SC}, and the low-energy many-body states will share a
nonvanishing expectation value $\langle \n_d-1\rangle$
[$=Q_{\loc}(T=0)$].
The bosons will couple to this impurity charge, yielding low-energy states
described most naturally in terms of displaced-oscillator states
[cf.\ Eq.\ \eqref{a:displaced}] annihilated by operators
\begin{equation}
\bar{a}_{\bq}
  = a_{\bq}
    + \frac{\lambda_{\bq}}{\sqrt{N_q} \, \omega_{\bq}} \langle \n_d-1\rangle .
\end{equation}
For $s<1$, the \eb\ coupling is relevant so $\lambda$ will scale to strong
coupling below a crossover temperature $T_L\ll T_{\K}$.

For $\delta_d\ne 0$ and very large values of $\lambda$, one instead
expects the bosons to localize the impurity at a high temperature
scale $T_L$ into a state with $\langle \n_d\rangle\simeq 0$ (for
$\delta_d>0$) or $\langle \n_d\rangle\simeq 2$ (for $\delta_d<0$).
For $T\lesssim T_L$, the impurity degrees of freedom will be frozen,
the bosonic spectrum will rapidly approach strong coupling, and the
conduction electrons will have an excitation spectrum corresponding to
$\H_{\FI}^{\NRG}$ in Eq.\ \eqref{H_FI} with a small value of $|V_0|$.

Given the equivalence of $\H_{\SC}^{\NRG}$ and $\H_{\FI}^{\NRG}$, it seems
likely that the low-energy behavior of the asymmetric model will be the same
in the small-$\lambda$ and large-$\lambda$ limits. This suggests that
the many-body eigenstates evolve adiabatically as the \eb\ coupling is
increased from $\lambda=0^+$ to $\lambda\to\infty$,
without the occurrence of an intervening QPT.

For $s=1$, the \eb\ coupling is marginal, rather than relevant. One again
expects a continuous evolution of the low-energy NRG spectrum with the bare
value of $\lambda$. However, in this Ohmic case, the bosonic excitations
should correspond to noninteracting displaced oscillators rather than the
(truncated) strong-coupling spectrum found for $0<s<1$.

\begin{figure}
\centering
\includegraphics[angle=270,width=0.8\columnwidth]{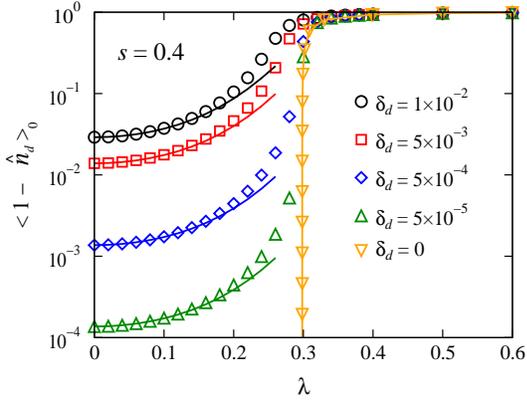}
\caption{\label{fig:1-nd}
(Color online)
Variation in the magnitude $\langle 1-\n_d\rangle_0$ of the ground-state
impurity charge with \eb\ coupling $\lambda$ for $s=0.4$, $U=0.1$,
$\Gamma=0.01$, $\Lambda=9$, $N_s=500$, and $N_b=8$. Symbols represent results
for five values of the impurity asymmetry $\delta_d = \epsilon_d + U/2$.
The solid lines corresponding to each case $\delta_d\ne 0$ represent the
impurity charge calculated by solving the Anderson model
[Eq.\ \protect\eqref{H_A}] for the same
$\delta_d$ value but using an effective Coulomb interaction 
$\tilde{U}^{\mathrm{NRG}}(0.3U)$ [Eq.\ \eqref{Ueff:NRG}]. The $\delta_d = 0$ symbols show values of
$\lim_{\phi\to 0^-} Q_{\loc}(\lambda,\phi;T=0)$, connected by an
interpolating line.}
\end{figure}

The preceding arguments are supported by our NRG results. Here, we illustrate
just the case $s=0.4$.
Figure \ref{fig:1-nd} shows the variation with $\lambda$ of the
ground-state expectation value $\langle 1-\n_d\rangle_0$ for several
values of $\delta_d$. In the symmetric case (the $\delta_d=0$ curve in
Fig.\ \ref{fig:1-nd}), the impurity charge vanishes throughout
the Kondo phase, and grows in power-law fashion on entry to the
localized phase. Away from particle-hole symmetry, by contrast,
$\langle 1-\n_d\rangle_0$ increases smoothly from its Anderson-model
value at $\lambda=0$ to approach 1 as $\lambda\to\infty$.

For all nonzero values of $\delta_d$, $\Gamma$ and $\lambda$, the low-energy
spectrum can be decomposed into the direct product of the fermionic spectrum
corresponding to $\H_{\SC}^{\NRG}(V_1)$ [or $\H_{\FI}^{\NRG}(V_0)$] and
the same localized-phase bosonic spectrum as found for the symmetric
model. The potential scattering $V_1$ (or $V_0$) is tied to
$\langle \n_d-1\rangle_0$ by Eq.\ \eqref{Friedel}, just as in the
Anderson model.

For small $\lambda$, the value of $\langle \n_d-1\rangle_0$ can be related to
the corresponding quantity in the Anderson model by making use of the effective
Coulomb interaction introduced in Sec.\ \ref{subsec:Gamma=0}. In the
asymmetric Anderson model, the ground-state charge becomes frozen once the
system passes out of its mixed-valence regime, i.e., somewhat below a
characteristic temperature $T_f$ defined\cite{Krishna-murthy:80} for
$\Gamma\ll -\epsilon_d\ll U$ as the solution of
\begin{equation}
\label{T_f}
T_f = |\epsilon_d| - \frac{\Gamma}{\pi} \ln \frac{U}{T_f} .
\end{equation}
In the charge-coupled BFA model, $U$ and $\epsilon_d$ in Eq.\ \eqref{T_f}
should presumably be replaced by $\tilde{U}(T_f)$ and $\delta_d - \frac{1}{2}
\tilde{U}(T_f)$, respectively. However, it suffices for our purposes to note that
$T_f$ can be expected to be of the same order as, but somewhat smaller than,
$|\epsilon_d|$. It is then reasonable to hypothesize that
$\langle n_d-1\rangle_0$ in the asymmetric charge-coupled BFA model should be
close to the ground-state impurity charge of the Anderson model with the same
$\Gamma$ and $\delta_d$, but with $U$ replaced by $\tilde{U}(E)$
[Eq.\ \eqref{Ueff:s>0}] evaluated at $E\simeq T_f$. Our numerical results
support this conjecture.
For example, Fig.\ \ref{fig:1-nd} shows that close to particle-hole symmetry
($\epsilon_d = -U/2$), the Anderson-model charge calculated for $\tilde{U}^{\mathrm{NRG}}(E)$ 
[Eq.\ \eqref{Ueff:NRG}] with $E=0.3U$ (solid lines) reproduces quite well the value of
$\langle \n_d - 1\rangle_0$ (symbols) over quite a broad range of
\eb\ couplings $0\le \lambda\lesssim\frac{2}{3}\lambda_c$, where
$\lambda_c\simeq 0.29835$ is the critical coupling of the symmetric problem.

\begin{figure}
\centering
\includegraphics[angle=270,width=0.8\columnwidth]{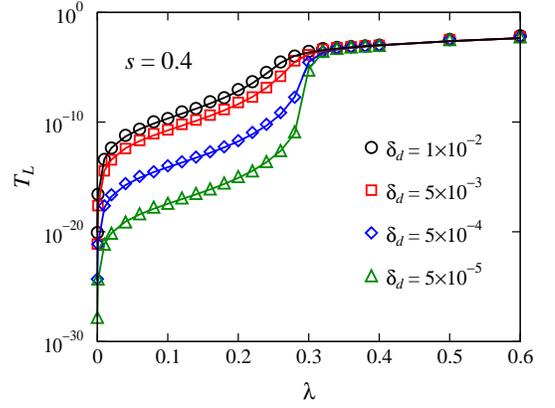}
\caption{\label{fig:T_L}
(Color online)
Variation in the bosonic localization temperature $T_L$
with coupling $\lambda$ for $s=0.4$, $U=0.1$, $\Gamma=0.01$,
$\Lambda=9$, $N_s=500$, $N_b=8$, and various impurity asymmetries
$\delta_d = \epsilon_d + U/2$. The solid lines were obtained by
evaluating Eq.\ \protect\eqref{T_L} with the $\langle 1-\n_d\rangle_0$
values shown in Fig.\ \protect\ref{fig:1-nd} and with $C_L=3$.}
\end{figure}

In the small-$\lambda$ limit, one can also estimate the boson-localization
temperature $T_L$ by considering the evolution with decreasing $T$ of the
effective value of $\lambda \langle \n_d - 1\rangle_0$. The impurity charge
does not renormalize, while to lowest order the effective \eb\ coupling
obeys\cite{Smith:99} Eq.\ \eqref{lambda:RG}.
Defining $T_L$ by the condition
$\tilde{\lambda}(T_L)|\langle\n_d - 1\rangle_0| = C_L$, we find
\begin{equation}
\label{T_L}
T_L \simeq \bigl| C_L^{-1} \lambda
  \langle \n_d - 1\rangle_0 \bigr|^{2/(1-s)} .
\end{equation}
In Fig. \ref{fig:T_L}, symbols represent $T_L$ values extracted from the
crossover of bosonic excitations in the NRG spectrum, while solid lines show the
results of evaluating Eq.\ \eqref{T_L} using $C_L=3$ and the
$\langle \n_d - 1\rangle_0$ values shown in Fig.\ \ref{fig:1-nd}.
The algebraic relation between the numerical values of $T_L$ and
$\langle 1-\n_d\rangle_0$ is well obeyed over a range of \eb\ couplings
that extends beyond $\lambda_c$ of the symmetric problem.

\begin{figure}
\centering
\includegraphics[angle=270,width=0.85\columnwidth]{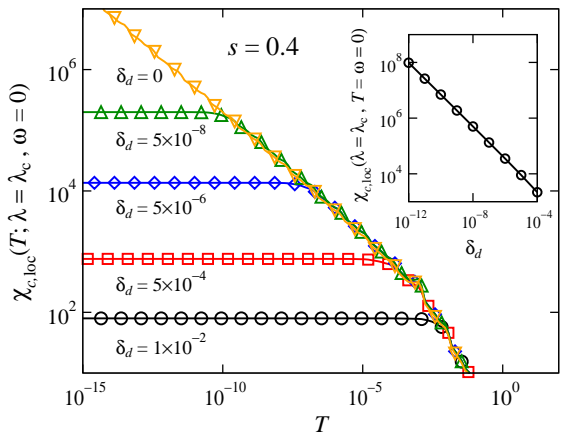}
\caption{\label{fig:static-chi:asymm}
(Color online)
Static local charge susceptibility $\chi_{c,\loc}(T;\omega=0)$ vs
temperature $T$ for $s=0.4$, $U=0.1$, $\Gamma=0.01$,
$\lambda\simeq 0.29835$, $\Lambda=9$, $N_s=500$, $N_b=8$,
and various impurity asymmetries $\delta_d = \epsilon_d + U/2$. The
\eb\ coupling equals the critical coupling $\lambda_c$ of the
symmetric case $\delta_d=0$. Inset: zero-temperature static local charge
susceptibility $\chi_{c,\loc}(\omega=T=0)$ vs $\delta_d$.}
\end{figure}

Figure \ref{fig:static-chi:asymm} plots the static local charge
susceptibility calculated for $s = 0.4$ at the critical \eb\ coupling of
the symmetric model. For $\delta_d\ne 0$, $\chi_{c,\loc}$ follows the
quantum critical behavior $\chi_{c,\loc}(T;\omega=0)\propto T^{-x}$ from a
high-temperature cutoff of order $T_{\K}$ down to a crossover temperature $T_*$,
below which the susceptibility saturates. Based on Eq.\ \eqref{ansatz} with
the identification $\phi\equiv\delta_d$, one expects
$T_*\propto|\delta_d|^{2/(1+x)}$ and, hence,
\begin{equation}
\label{chi:s=1}
\chi_{c,\loc}(\phi;\lambda=\lambda_c,\omega=T=0) \propto|\delta_d|^{-2x/(1+x)}.
\end{equation}
The log-log plot in the inset of Fig.\ \ref{fig:static-chi:asymm} has a slope
0.57 that is fully consistent with Eq.\ \eqref{chi:s=1}.

The results of this work show that gaining direct access to the quantum
critical point of the charge-coupled BFA model requires simultaneous fine
tuning of two parameters: the \eb\ coupling $\lambda$ as a function of
the hybridization $\Gamma$ and the on-site Coulomb repulsion $U$; and the
particle-hole asymmetry (determined in our calculations solely by
$\delta_d=\epsilon_d+U/2$, but in general also affected by the shape of
the conduction-band density of states).
While it may prove very challenging, or even impossible, to achieve this
feat in any experimental realization of the model, it should be a more
feasible task to carry out a rough tuning of parameters that places the
system in the quantum critical regime over some window of elevated
temperatures and/or frequencies.

\section{Summary}
\label{sec:summary}

We have conducted a detailed study of the charge-coupled Bose-Fermi
Anderson model, in which a magnetic impurity both hybridizes with a
structureless conduction band and is coupled, via its charge, to a
dissipative environment represented by a bosonic bath having a spectral
function that vanishes as $\omega^s$ for vanishing frequencies
$\omega\to 0$. With increasing coupling between the impurity and the
bath, we find a crossover from a conventional Kondo effect---involving
conduction-band screening of the impurity spin degree of freedom---to a
charge-Kondo regime in which the delocalized electrons quench impurity
charge fluctuations.

Under conditions of strict particle-hole symmetry, further increase in the
impurity-bath coupling gives rise for $0<s\le 1$ to a quantum phase transition
between the Kondo phase, in which the static charge and spin susceptibilities
approach constant values at low temperatures, and a localized phase in which
the static charge susceptibility exhibits a Curie-Weiss behavior indicative
of an unquenched local charge degree of freedom. For sub-Ohmic bosonic bath
spectra (described by an exponent $s$ satisfying $0<s<1$), the continuous
quantum phase transition is governed by an interacting critical point
characterized by hyperscaling relations of critical exponents and $\omega/T$
scaling in the dynamical local charge susceptibility. Moreover, the continuous
phase transition of the present model belongs to the same universality class
as the transitions of the spin-boson and the Ising-anisotropic Bose-Fermi
Kondo models. For an Ohmic ($s=1$) bosonic bath spectrum, the quantum phase
transition is of Kosterlitz-Thouless type.

In the presence of particle-hole asymmetry, the quantum phase transition
described in the previous paragraph is replaced by a smooth crossover, but
for small-to-moderate asymmetries, signatures of the symmetric quantum critical
point remain in the physical properties at elevated temperatures and/or frequencies.
Investigation of the regime of strong particle-hole asymmetry, and of
self-consistent versions of the charge-coupled Bose-Fermi Anderson model that
arise with the extended dynamical mean-field theory of lattice fermions, will be
pursued in future work.

\begin{acknowledgments}
We thank Brian Lane for useful discussions. Much of the computational work
was performed at the University of Florida High-Performance Computing Center.
This work was supported in part by NSF Grants No.\ DMR-0710540 (M.C.\ and K.I.)
and No.\ DMR-0706625 (M.T.G.).
\end{acknowledgments}


\begin{thebibliography}{10}
\bibitem{large-Gammas}
A number of the results presented in Sec.\ \ref{sec:sub-ohmic} were
obtained using unphysically large values of the hybridization $\Gamma$.
These values were employed to accelerate the convergence of the NRG
levels to the critical spectrum, and thereby to minimize computational
rounding errors.
\bibitem{critical_decomposition}
This decomposition of the BFK critical spectrum was not explicitly noted
in Refs.\ \protect\onlinecite{Glossop:05} and \protect\onlinecite{Glossop:07}.
However, it can be understood (following arguments analogous to those
presented here for the charge-coupled BFA model) under the assumption that,
near the phase boundary, the longitudinal exchange coupling renormalizes
rapidly to $J_z=\infty$.
\bibitem{tracking}
From Eqs.\ \protect\eqref{nu:defn} and \protect\eqref{gamma:defn}, and
Tables \protect\ref{tab:nu} and \ref{tab:exponents}, one expects the
vanishing of $\Gamma_{\K}\propto T_*$ and $4/\chi_{c,\loc}(0,0)$ on approach
to the critical point to be governed by different powers of
$\lambda_c - \lambda$.
\end{thebibliography}
\end{document}